\documentclass{article}
\usepackage[textwidth=17cm]{geometry}
\usepackage[utf8]{inputenc}
\usepackage{amsmath,amssymb,bm}
\usepackage{mathtools}
\usepackage{graphicx}
\usepackage[dvipsnames]{xcolor}
\usepackage{dirtytalk}
\usepackage[colorlinks,linkcolor=blue,citecolor=magenta]{hyperref}
\usepackage{doi}
\usepackage[nameinlink]{cleveref}

\usepackage{authblk}

\usepackage[linesnumbered,vlined,algoruled]{algorithm2e}
\crefname{algocf}{alg.}{algs.}
\Crefname{algocf}{Algorithm}{Algorithms}

\newcommand{\minimize}{\operatornamewithlimits{minimize}}

\usepackage{bbm}

\newcommand{\vpar}{v_{\parallel}}

\newcommand{\vperp}{v_{\perp}}
\newcommand{\vmax}{v_{\max}}

\newcommand{\nfp}{n_{\text{fp}}}
\newcommand{\tmax}{t_{\max}}
\newcommand{\R}{\mathbb{R}}
\newcommand{\Ebb}{\mathbb{E}}

\newcommand{\Bb}{\mathbf{B}}

\newcommand{\wb}{\mathbf{w}}
\newcommand{\xb}{\mathbf{x}}

\newcommand{\Jc}{\mathcal{J}}
\newcommand{\Pc}{\mathcal{P}}

\newcommand{\Tc}{\mathcal{T}}

\newcommand{\Jenergy}{\mathcal{J}_{\text{energy}}}

\newcommand{\nw}{{n_{\wb}}}
\newcommand{\nmode}{n_{\text{mode}}}

\newcommand{\ns}{{n_{s}}}
\newcommand{\ntheta}{{n_{\theta}}}
\newcommand{\nzeta}{{n_{\zeta}}}
\newcommand{\ngrid}{{n_{\text{grid}}}}

\def\code#1{\texttt{#1}}
\newcommand{\response}[1]{\textcolor{black}{#1}}

\title{Direct Optimization of Fast-Ion Confinement in Stellarators}
\author[1]{David Bindel}
\author[2]{Matt Landreman}
\author[3,*]{Misha Padidar}

\affil[1]{Department of Computer Science, Cornell University, Ithaca, NY 14850, USA}
\affil[2]{Institute for Research in Electronics and Applied Physics, University of Maryland, College Park, MD 20742, USA}
\affil[3,*]{Center for Applied Mathematics, Cornell University, Ithaca, NY 14850, USA}
\affil[*]{Corresponding author: Misha Padidar, map454@cornell.edu}
\date{}

\begin{document}

\maketitle

\begin{abstract}
     Confining energetic ions such as alpha particles is a prime concern in the design of stellarators. However, directly measuring alpha confinement through numerical simulation of guiding-center trajectories has been considered to be too computationally expensive and noisy to include in the design loop, and instead has been most often used only as a tool to assess stellarator designs post hoc. In its place, proxy metrics, simplified measures of confinement, have often been used to design configurations because they are computationally more tractable and have been shown to be effective. 
     % Despite the success of proxies, it is unclear what is being sacrificed by using them to design the device rather than relying on direct trajectory calculations. 
     Despite the success of proxies, \response{their correlation with direct trajectory calculations is known to be imperfect}.
     In this study, we optimize stellarator designs for improved alpha particle confinement without the use of proxy metrics. 
     % In particular, within an optimization loop we measure alpha particle losses by numerically simulating alpha particle trajectories. 
     In particular, we numerically optimize an objective function that measures alpha particle losses by simulating alpha particle trajectories.
     While this method is computationally expensive, we find that it can be used successfully to generate configurations with low losses.
\end{abstract}

\section{Introduction}
\label{sec:introduction}

Alpha particles are born in stellarators as a product of the fusion reaction.
Born with $3.5$ MeV, alpha particles carry a substantial amount of energy which, if confined, will heat the plasma and sustain the reaction.
On the other hand, poor confinement of the alphas can have destructive effects on the plasma-facing components, and detract from plasma self-heating.
Hence, confinement of fast ions is, and has been, a focal point in stellarator design
\cite{ku2008physics,henneberg2019properties,bader2019stellarator,landremanBullerDrevlak,leviness2022energetic}.

Stellarator design is generally split into two stages. In the first stage the plasma shape is optimized such that the magnetohydrodynamic (MHD) equilibrium meets specified performance criteria, such as particle confinement, stability, and/or reduced turbulence. The second stage is then devoted to finding electromagnetic coil shapes and currents which generate the desired magnetic field. Due to the computational expense of simulating particle trajectories for long times, typically stage-one configurations are designed using proxy metrics for confinement, such as quasisymmetry (QS) \cite{boozer1983transport,landreman2022magnetic,rodriguez2022measures},  $\Gamma_c$ \cite{nemov2008poloidal,bader2021modeling,leviness2022energetic, bader2019stellarator}, \response{$\Gamma_\alpha$ \cite{velasco2021model}}, and epsilon effective, $\epsilon_{eff}$ \cite{nemov1999evaluation}. Recently, numerical optimization of a QS metric has been particularly successful in improving particle confinement in stellarators, leading to configurations with less than $1\%$ fast-ion losses \cite{landreman2022magnetic,wechsung2022precise,landremanBullerDrevlak}. Despite the success of QS and other proxies,  
%it is unclear what is being sacrificed by using proxies to design the device rather than relying on exact calculations. 
\response{their correlation with direct trajectory calculations is known to be imperfect \cite{bader2021modeling, paul2022energetic, leviness2022energetic}.}
%For example, s
Since QS is a sufficient condition for confinement, rather than a necessary condition, it may be overly stringent. 
\response{
The quantity $\epsilon_{eff}$ provides a rigorous measure of neoclassical transport for the thermal plasma in a certain regime of collisionality, but it neglects the poloidal drift that is known to be important for energetic particles.
The figures of merit $\Gamma_c$ and $\Gamma_\alpha$ do account for this effect, and have been constructed specifically as surrogates for confinement of energetic particles.
These quantities are far faster to compute than direct calculation of an ensemble of guiding center trajectories, and have been used successfully in optimization \cite{bader2019stellarator, leviness2022energetic, sanchez2022quasi}.
Nonetheless $\Gamma_c$ and $\Gamma_\alpha$ do not directly account for several effects such as the finite width of trapped orbits, resonances of the bounce motion \cite{park2009nonambipolar, white2022poor, albert2022resonant}, diffusion of banana orbits \cite{goldston1981confinement, white1996ripple}, and diffusion associated with trapping/detrapping transitions \cite{beidler2001stochastic,tykhyy2021theory}.
Other measures of energetic particle confinement based on orbit classification are also being investigated \cite{albert2020accelerated}.
}
%Similarly, proxies in general only approximate the true goal of improving particle confinement, and do not capture the goal holistically or exactly.

In this study we opt for a direct approach to achieve fast-ion confinement: we optimize stellarator designs by simulating fast-ion trajectories and minimizing the empirical loss of energy. Our model takes the form
\begin{equation}
\begin{split}
    \minimize_{\wb \in \R^{\nw}} \quad &\Jc(\wb) := \Ebb_{\xb,\vpar}[\Jenergy(\xb,\vpar,\wb)]
    \\
    % & c_i(\wb) \le 0 \quad i=1,\ldots,\nsc
    & B_{-}^* \le B(\xb,\wb) \le B_+^* \quad \forall \xb \in \Pc
\end{split}
\label{eq:main}
\end{equation}
The objective $\Jc$ measures the expected value of the energy lost, $\Jenergy$, due to alpha particles born with a random initial position, $\xb$, and parallel velocity, $\vpar$, drifting through the last closed flux surface of the plasma. The decision variables $\wb \in\R^{\nw}$ are Fourier coefficients representing the shape of the plasma boundary. Motivated by physical and engineering requirements, the infinite dimensional nonlinear bound constraints restrict the strength of the magnetic field $B$ to an interval $[B_{-}^*,B_{+}^*]$ at each point throughout the plasma volume $\xb\in\Pc$. 

By varying the shape of the plasma boundary we seek MHD equilibria that minimize the loss of alpha particle energy. 
% The expected energy lost, $\Jc(\wb)$, is computed empirically by randomly generating alpha particles according to a birth distribution, and simulating their trajectories in Boozer coordinates via the collision-less guiding center equations. 
 The expected energy lost, $\Jc(\wb)$, is computed empirically from Monte Carlo simulation of collision-less guiding center trajectories by use of an approximation for the alpha particle energy in terms of confinement time. Due to the lack of analytical derivatives, we solve \cref{eq:main} using derivative-free optimization methods. In this document, we discuss practical challenges such as the noisy objective computation, high computational cost, and choice of derivative-free optimization algorithm. Our numerical results show that the approach is indeed effective at finding desirable configurations, and that the configurations we find are visibly not quasi-symmetric. 

To the author's knowledge, only two stellarators have been designed by simulating alpha particle losses within the design loop: the ARIES-CS stellarator \cite{ku2008physics} and a design by Gori et. al. \cite{gori2001alpha}. In the design of ARIES-CS, the average confinement time of $\sim 500$ particles was included as a term in an optimization objective. The initial particle locations were held fixed during the optimization, leading to an \say{effective and robust} technique. Similarly, Gori et al. included the average confinement time of reflected particles in their optimization objective. To mitigate the high computational cost and time required to simulate particle trajectories both studies limited the particle simulation to a fixed number of toroidal transits. Despite the empirical success of these designs, there is not a clear description of the methods used and challenges faced. As part of our work we bring light to this approach.

\response{While the alpha confinement in ARIES-CS was improved compared to the National Compact Stellarator eXperiment (NCSX) \cite{zarnstorff2001physics}, which was used as the initial condition for the ARIES-CS optimization, alpha losses in ARIES-CS are in fact still rather large \cite{bader2021modeling,landreman2022magnetic}. 
This may be due in part to the small number of particles and short integration times that were feasible at the time, and to other conflicting objectives such as coil feasibility and MHD stability.
Evidently, direct use of particle trajectories in the objective function is no guarantee that good confinement will result. Hence, another goal of the present work is to see how much further the alpha confinement can be optimized when focusing on only this objective, and with more particles to make the objective smoother.}

The paper is structured as follows. In \Cref{sec:physics} we discuss the life cycle of alpha particles and the relevant physics to our numerical simulations. \Cref{sec:simsopt} describes the computational workflow for modeling and evaluating candidate stage-one designs. In \Cref{sec:optimization_model} we mathematically formulate our design problem as an optimization problem. \Cref{sec:objective_approximation} compares methods of computing the objective function via the simulation of alpha particle trajectories. Numerical results are presented in \Cref{sec:numerical}, prior to a brief discussion of future research directions in \Cref{sec:conclusion}.

% \paragraph{Notation}
% \todo{delete notation section, and redistribute notation across the paper.}
% $\Bb\in\R^3$ is the magnetic field, with field strength $B = |\Bb|$. $\vpar,\vperp$ are the parallel and perpendicular velocity to field lines, $\mu$ is the magnetic moment and $\xb = (s,\theta,\zeta)$ are Boozer coordinates. The toroidal flux is $\Psi_T$. $f$ denotes a probability distribution. The temperature profile is $\Tc(s)$, and the number densities for tritium and deuterium are $n_T,n_D$. $\tmax$ is the maximum trace time used in tracing particles numerically. We denote the confinement time, within a terminal time $\tmax$, of a particle with initial position $(\xb,\vpar)$ as $T(\xb,\vpar)$, or more simply $T$ for notational convenience. Expectation, in the statistical sense, is denoted $\Ebb$. The number of field periods is $\nfp$. $\Pc$ denotes the plasma volume enclosed by the last closed flux surface, and $[-\vmax,\vmax]$ are bounds on the fast ion speed. A superscript $*$ denotes user specified target values, such as the target aspect ratio $A^*$, target rotational transform $\iota^*$, and target mean field strength $B^*$, field strength bounds $B_-^*,B_+^*$, target minor radius $a^*$, and target mirror ratio $r^*$. $\wb \in\R^\nw$ are the decision variables. 
% $\nsc$ denotes the number of simulation based constraints used in the optimization.

\section{Physical Model}
\label{sec:physics}

We consider toroidal plasma configurations that are static MHD equilibria, satisfying $\mu_0^{-1}(\nabla\times\Bb)\times\Bb=\nabla p$, where $p$ is the pressure and $\Bb\in\R^3$ is the magnetic field.
It is assumed that nested toroidal flux surfaces exist.
For the numerical experiments in this work, we adopt the low $\beta$ (plasma pressure divided by magnetic pressure) limit of $p\approx 0$ and $\nabla\times\Bb \approx 0$ for simplicity, but the methods here are fully applicable to MHD equilibria with substantial pressure and current.
A convenient coordinate system for MHD equilibria is Boozer coordinates $\xb = (s,\theta,\zeta)$, where $s$ is the toroidal flux normalized to be 1 at the plasma boundary, and $\theta$ and $\zeta$ are poloidal and toroidal angles. The domain of the coordinates is $s,\theta,\zeta \in \Pc := [0,1]\times [0,2\pi) \times [0,2\pi/\nfp)$ for a stellarator with $\nfp$ field periods. 
% For convenience we collect the Boozer coordinates into the vector $\xb = (s,\theta,\zeta)$.

Motion of alpha particles in the equilibrium is modeled using the collisionless guiding center equations. For the case considered here of low $\beta$, these equations are
\begin{align}
\begin{split}
    % \frac{d\mathbf{R}}{dt} &= \vpar\bb 
    % + \frac{m}{qB^3}\left(\vpar^2+ \frac{\vperp^2}{2}\right)\Bb\times \nabla B,
    % \\
    % \frac{d\vpar}{dt} &= -\frac{\vperp^2}{2B}\bb\cdot\nabla B.
    \frac{ds}{dt} &= - \frac{\partial B}{\partial\theta} \frac{m}{q\psi_a} \left(\frac{\vpar^2}{B} + \mu \right),
    \\
    \frac{d\theta}{dt} &= \frac{\partial B}{\partial s} \frac{m}{q\psi_a} \left(\frac{\vpar^2}{B} + \mu \right) + \frac{\iota\vpar B}{G},
    \\
    \frac{d\zeta}{dt} &=  \frac{\vpar B}{G},
    \\
    \frac{d\vpar}{dt} &=  -\left(\iota\frac{\partial B}{\partial\theta} + \frac{\partial B}{\partial\zeta} \right) \frac{\mu B}{G}.
\end{split}
\label{eq:gc_motion_cart}
\end{align}
Here, $t$ is time, $m$ is the particle's mass, $q$ is the particle's charge, $B=|\Bb|$ is the field strength, $\mathbf{b}=\Bb/B$, \response{$2\pi\psi_a$ is the toroidal flux at the plasma boundary, $G$ is 
the poloidal current outside the flux surface times $\mu_0/(2\pi)$,} and $\vpar$ and $\vperp$ are the components of velocity parallel and perpendicular to $\Bb$. \response{This representation of the guiding center motion can be derived by substituting $\Bb = G\nabla \zeta$ into the vacuum guiding center equations with curvature and $\nabla B$-drifts.}
The magnetic moment $\mu = \vperp^2/(2B)$ is conserved, as is the speed $v=\sqrt{\vpar^2+\vperp^2}$. Trapped particles, which have sufficiently small $|\vpar/\vperp|$, experience  reversals in the sign of $\vpar$.
Particles that do not experience $\vpar$ sign reversals are called ``passing".

Alpha particles are born isotropically with an energy of 3.5 MeV.
We consider two models for the initial spatial distribution.
The first model is based on the local fusion reaction rate, resulting in alpha particle birth throughout the plasma volume.
% In the second initial spatial distribution considered, alpha particles are born only on a specified flux surface.
The second model distributes alpha particles across a single specified flux surface.
Either way, after birth, alpha particle guiding centers are followed for a specified amount of time, or until they exit the plasma boundary surface, at which time they are considered lost.

\begin{figure}[tbh!]
\includegraphics[scale=0.5]{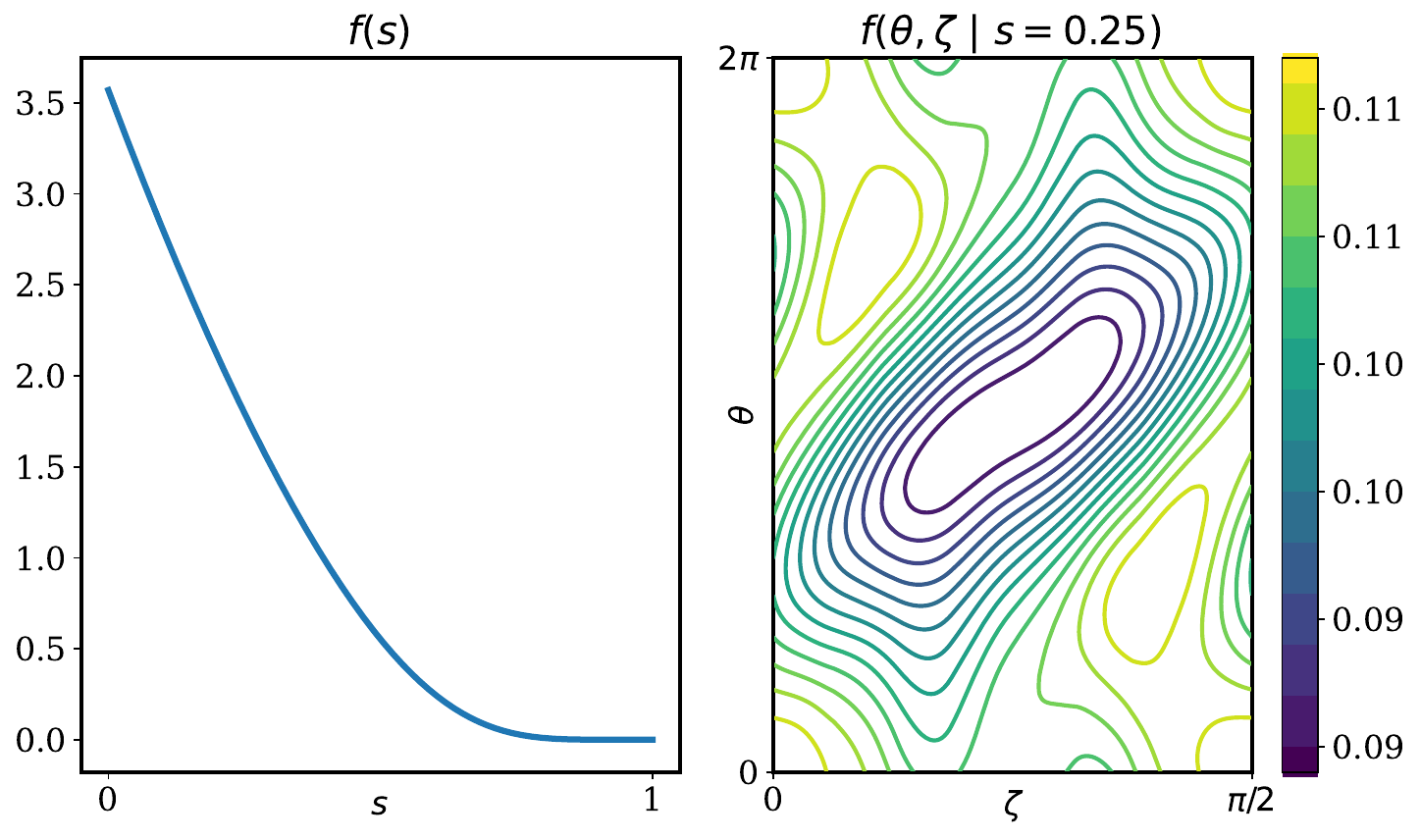}
\centering
\caption{(left) Radial probability density $f_s(s)$ derived from the fusion reaction rate. (right) Density over Boozer coordinates $\theta$ and $\zeta$, $f_{\theta,\zeta}$ for configuration \textbf{A} which will be discussed in \Cref{sec:numerical}.}
\label{fig:sample_density}
\end{figure}

% Alpha particles are initialized in a standard manner \cite{drevlak2014, bader2021modeling,landremanBullerDrevlak,leviness2022energetic}, as follows.
The birth distribution of alpha particles is derived in a standard manner \cite{drevlak2014, bader2021modeling,landremanBullerDrevlak,leviness2022energetic}, as follows.
For calculations in which alpha particles are born throughout the volume, the spatial birth distribution is proportional to the local reaction rate \cite{NRLPlasma2019}, $f(s,\theta,\zeta) \propto n_D n_T  (\overline{\sigma v})_{DT}$. 
Here, $D$ and $T$ subscripts indicate deuterium and tritium, $n_D$ and $n_T$ are the species densities, which we assume to be equal, and $(\overline{\sigma v})_{DT}$ is the Maxwellian-averaged fusion cross-section, computed in \cite{NRLPlasma2019} by
%The Maxwellian-averaged cross-section is computed by \cite{NRLPlasma2019}
\begin{equation}
    (\overline{\sigma v})_{DT}(s) = 3.68 \times 10^{-18}T_i^{-2/3}(s)\exp(-19.94T_i^{-1/3}(s)) \;\text{m}^3\text{sec}^{-1},
\end{equation}
where $T_i$ is the ion temperature in keV.
Within the numerical experiments, we assume the following density and temperature profiles:
\begin{align}
    n_D(s) &= n_T(s) = 2\times 10^{20}(1 - s^5) \; \text{m}^{-3},
    \\
    T_i(s) &= 12(1 - s) \; \text{keV}.
\end{align}
%where $s$ is the toroidal flux normalized to be 1 at the plasma boundary.
These density and temperature profiles reflect plausible reactor parameters \cite{ku2008physics,alonso2022physics}, and the fact that temperature profiles in experiments are typically more peaked than density profiles.
In this study the temperature and density profiles are held fixed in order to focus on the optimization of particle trajectories.
The radial birth distribution of particles is thus proportional to 
\begin{equation}
    f_s(s) \propto (1 - s^5)^2 (1 - s)^{-2/3}\exp(-19.94 (12(1 - s))^{-1/3}).
    \label{eq:birth_distribution_radial}
\end{equation}
as depicted in \Cref{fig:sample_density} (left). Alternatively, to only consider particles born on a single flux surface, the localized initial radial distribution can be expressed as $f_s(s) = \delta(s - s_0)$, where $s_0=0.25$ is used in numerical experiments.

For either initial radial distribution, particles are initialized uniformly over flux surfaces. This uniformity is expressed by a determinant of the Jacobian from Boozer to Cartesian coordinates $\sqrt{g}$,
\begin{equation}
    f_{\theta,\zeta}(\theta,\zeta \, | \, s) \propto |\sqrt{g}| .
    \label{eq:birth_distribution_angles}
\end{equation} 
 \Cref{fig:sample_density} (right) shows $f_{\theta,\zeta}$ for configuration \textbf{A}, which will be discussed in \Cref{sec:numerical}.
Lastly, the isotropic velocity birth distribution corresponds to a uniform distribution of $\vpar$ over $[-\vmax,\vmax]$, where $\vmax=\sqrt{2E/m}$ and $E=3.5$ MeV.
Defining the associated distribution
\begin{equation}
    f_{\vpar}(\vpar) = \frac{1}{2\vmax},
    \label{eq:birth_distribution_vpar}
\end{equation}
the total birth distribution is
\begin{equation}
    f(s,\theta,\zeta,\vpar) = f_s(s)f_{\theta,\zeta}(\theta,\zeta \, | \, s)f_{\vpar}(\vpar).
    \label{eq:birth_distribution}
\end{equation}

Several mechanisms exist by which trapped particles are lost \cite{gibson1967single,galeev1969plasma,goldston1981confinement,beidler2001stochastic,paul2022energetic}.
``Ripple trapped'' particles, those trapped in a single field period or in coil ripple, typically experience a nonzero average radial magnetic drift and so are quickly lost.
Other trapped particle trajectories may resemble the banana orbits of a tokamak, but with radial diffusion due to imperfect symmetry.
Particles that transition between these two types of trapped states make additional radial excursions.
Particles with wide banana orbits may also be directly lost.
Generally, passing particles are not lost unless they are born very close to the plasma boundary.

\section{Modeling and optimization software}
\label{sec:simsopt}
To evaluate candidate stage-one stellarator designs we rely on the \code{SIMSOPT} code \cite{landreman2021simsopt}. \code{SIMSOPT} is a framework for stellarator modeling and optimization which interfaces with MHD equilibrium solvers such as \code{VMEC} \cite{hirshman1986three} and \code{SPEC} \cite{hudson2012computation}, and houses infrastructure for defining magnetic fields, computing coordinate transformations, tracing particles, and computing properties of fields and equilibria. Certain rate-limiting computations in \code{SIMSOPT}, such as evaluating magnetic fields, are executed in C++. For ease of use, however, Python bindings are used through the PyBind11 library, allowing users to interface with \code{SIMSOPT} solely through the Python interface. 

In order to design stage-one configurations we first find an ideal MHD equilibrium by evaluating \code{VMEC} with a prescribed plasma boundary shape, current profile, and pressure profile. Subsequently, the magnetic field is transformed to Boozer coordinates which is used within the guiding center equations when tracing particles. 
% In our numerical studies we only consider vacuum configurations, hence the current and pressure profiles are identically zero. 

The plasma boundary is paramterized as a Fourier series in the poloidal and toroidal angles $\theta$ and $\phi$,
\begin{equation}
\begin{split} 
    R(\theta,\phi) &= \sum_{n=0}^{\nmode} R_{0,n}\cos(- n_{\text{fp}}n\phi) + \sum_{m =1}^{\nmode} \sum_{n=-\nmode}^{\nmode} R_{m,n}\cos(m\theta - n_{\text{fp}}n\phi),
    % R(\theta,\phi) &= \sum_{m =0}^{\nmode} \sum_{n=-\nmode}^{\nmode} R_{m,n}\cos(m\theta - n_{\text{fp}}n\phi),
    \\
    Z(\theta,\phi) &= \sum_{n=1}^{\nmode} Z_{0,n}\sin(- n_{\text{fp}}n\phi)  + \sum_{m =1}^{\nmode} \sum_{n=-\nmode}^{\nmode} Z_{m,n}\sin(m\theta - n_{\text{fp}}n\phi),
    % Z(\theta,\phi) &= \sum_{m =0}^{\nmode} \sum_{n=-\nmode}^{\nmode} Z_{m,n}\sin(m\theta - n_{\text{fp}}n\phi).
\end{split}
    \label{eq:vmec_fourier_rep}
\end{equation}
where $\nmode = \{0,1,2,\ldots\}$ can be increased to achieve more complicated boundary representations. Field period symmetry with $\nfp$ periods and stellarator symmetry have been assumed. \response{While a separate value of $\nmode$ could be used for the toroidal and poloidal modes, we follow \cite{landreman2022magnetic} in using a single value for simplicity.} %Note that for $m=0$ we skip the term for the $n<0$ cosine terms, and the $n\le 0$ sine terms 

Upon computing the equilibrium, a Boozer-coordinate representation of the magnetic field is computed using the \code{BoozXform} code, via \code{SIMSOPT}. Working in Boozer coordinates reduces the number of interpolations required to integrate the guiding center equations. Initial particle positions and parallel velocities can then be generated, and particles traced using the vacuum guiding center equations in Boozer coordinates up to a terminal time $\tmax$ or until stopping criteria are satisfied. The guiding center equations are solved using the adaptive Runge-Kutta scheme RK45. 
\begin{figure}[tbh!]
\includegraphics[scale=0.35]{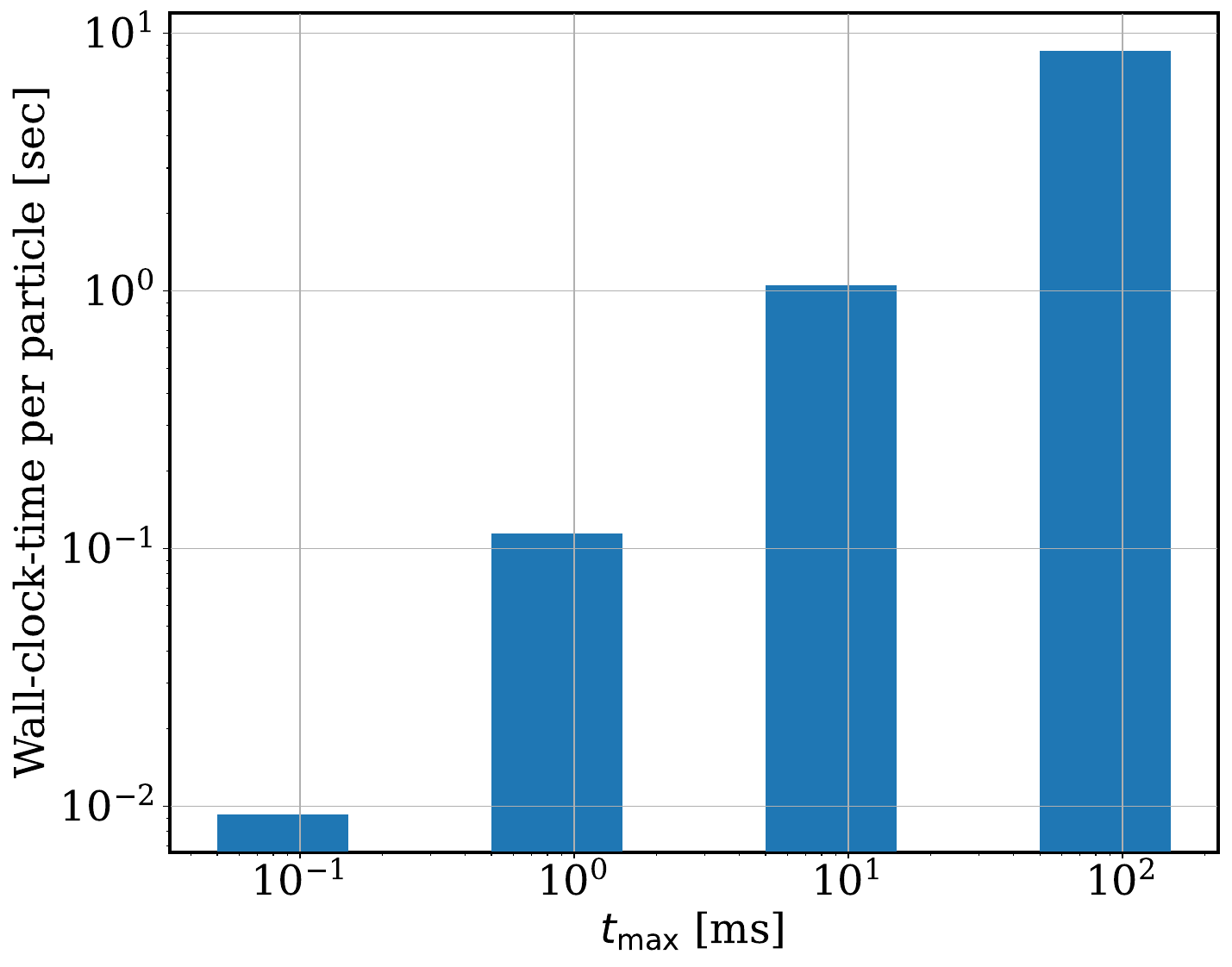}
\centering
\caption{Wall-clock time required to trace a single particle until a terminal time $\tmax$ using a single \response{Intel(R) Xeon(R) CPU E5-2620 v3 @ 2.40GHz} processor on a computing cluster. \response{The processor was allotted 2GB of memory}. Timing results were averaged over $2000$ particles randomly generated throughout a four field period configuration, all of which were confined to their terminal time $\tmax$. The total time of an objective evaluation also includes the fixed time of evaluating VMEC, computing the Boozer transformation, and building interpolants of the $\Bb$-field, which took $19.07$ seconds for this configuration.
}
\label{fig:timing}
\end{figure}

VMEC and the particle tracing codes allow parallelism through MPI. While VMEC can be run efficiently on a single core, particle tracing is embarrassingly parallel and benefits from the use of numerous cores. Even with dozens of MPI processes, particle tracing can take anywhere from seconds to minutes of wall-clock time to complete. In addition, there are substantial costs, typically around $\sim 20$ seconds, associated with running VMEC, computing the Boozer transform, and interpolating the fields required for tracing. Timing results for simulating particle orbits are shown in \Cref{fig:timing}. In total, computing an equilibrium and tracing enough particles to evaluate the objective function, defined in \Cref{sec:objective}, often takes between 30sec and 130sec of wall-clock-time, depending on the terminal trace time, the configuration, and the number of particles. The optimization process, which can be run on a single node, or multiple nodes on a computing cluster, is time consuming, often running for one to two days. For example, solving an optimization problem would consume $26$ hours of wall-clock-time when using 48 MPI processes and a computational budget of 1000 function evaluations which each require tracing 3500 particles to $10$ms. This poses a serious challenge in performing the optimization. In \Cref{sec:conclusion} we discuss future work that could reduce this burden.

\section{Optimization model formulation}
\label{sec:optimization_model}
We now outline a mathematical optimization problem that seeks stellarator configurations with good confinement of fast-ions. By varying the shape of the plasma boundary we minimize the energy lost due to alpha particles exiting the last closed flux surface. In the following, we describe the salient characteristics of the problem: the representation of decision variables, nonlinear constraints on the magnetic field strength, and an objective that quantifies the confinement of alpha particle energy.

\subsection{Decision variables}
\label{sec:decision_variables}
The independent decision variables for optimization are the Fourier coefficients $R_{m,n},Z_{m,n}$ which define the shape of plasma boundary in \code{VMEC} via \cref{eq:vmec_fourier_rep}. The number of modes used in the boundary description is controlled by the parameter $\nmode\in\{0,1,2,\ldots\}$. Increasing $\nmode$ increases the complexity of the boundary shape allowing for potential improvements in confinement, while setting $\nmode=0$ only allows the major radius $R_{0,0}$ to vary. The total number of decision variables satisfies $\nw = 4\nmode^2 + 4\nmode$.

The major radius of a design is central to particle trajectories simulation, since the Larmour radius and guiding center drifts scale with the square of the ratio of major radius to aspect ratio, $\propto (R_{0,0}/A)^2$. Standardization of the device size is thus necessary in order to have realistic particle losses, and to prevent the optimization from shrinking the aspect ratio arbitrarily. In confinement studies, device size is typically standardized by constraining the minor radius or the plasma volume. We opt to constrain the minor radius \textit{implicitly} to $a \approx a^* := 1.7$m (the minor radius of ARIES-CS), by fixing the major radius, fixing the toroidal flux, and constraining the field strength. In particular, we fix the major radius based on the target aspect ratio $A^*:=7$,
\begin{equation}
    R_{0,0} = a^*A^*.
    \label{eq:major_radius_constraint}
\end{equation}
In \Cref{sec:constraints}, the toroidal flux and mean field strength will be selected to encourage the design to have an aspect ratio near $A^*$. If the design achieves the aspect ratio of $A^*$, it would also have an average minor radius of $a^*$. Otherwise, the minor radius will only be near $a^*$.
The decision variables are collected into the vector $\wb\in\R^{\nw}$ via $\wb = (R_{0,1},\ldots, Z_{0,0},\ldots)$.

\subsection{Nonlinear constraints}
\label{sec:constraints}

Engineering limitations on electromagnetic coils and the associated support structure place an upper limit on the magnetic field strength. For low-temperature superconductors, the field strength is limited to be no more than $15$T in the coil and approximately $5$T throughout the plasma volume \cite{ku2008physics}. To achieve reactor relevant scaling of the magnetic field, we fix the toroidal flux so that if the plasma has an average minor radius of $a^*$, the volume-averaged magnetic field strength is $B^*:=5$T,
\begin{equation}
    2\pi\psi_a = \pi (a^*)^2B^*.
    \label{eq:toroidal_flux_constraint}
\end{equation}
The value of toroidal flux set in \cref{eq:toroidal_flux_constraint} is used as an input parameter to the MHD equilibrium calculations, and does not need to be treated as a constraint in the optimization. When paired with the major radius constraint, \cref{eq:major_radius_constraint}, the toroidal flux constraint, \cref{eq:toroidal_flux_constraint}, to zeroth order fixes the ratio of the the squared aspect ratio to volume-averaged magnetic field strength, i.e. $A^2/B \approx (A^*)^2/B^*$. Thus by placing bound constraints on the field strength we can constrain the range of the aspect ratio. 

In addition, bound constraints on the field strength are necessary in order to constrain the mirror ratio $\max_{\xb}{B(\xb)}/\min_{\xb}{B(\xb)}$, which we find increases to unphysically large values when left unconstrained in optimization. We globally bound the field strength,
\begin{equation}
    B_{-}^* \le B(\xb) \le B_{+}^* \quad \forall \, \xb \in \Pc.
    \label{eq:b_field_constraint_infinite}
\end{equation}
%The upper and lower bounds $B^*_+ = B^*(1+\epsilon_B)$ and $B^*_- = B^*(1-\epsilon_B)$ with $\epsilon_B = (r^*-1)/(r^*+1)$.
The upper and lower bounds $B^*_+ = B^*\frac{2r^*}{r^*+1}$ and $B^*_- = B^*\frac{2}{r^*+1}$ enforce that the mirror ratio is at most $r^* :=1.35$, similar to W7-X and the Compact Helical System (CHS) \cite{beidler2011benchmarking, nishimura1990compact}. The upper bound on the field strength is derived from material properties and tolerances in coil engineering and the lower bound is motivated by requirements on confinement and transport based phenomena. The constraints \cref{eq:b_field_constraint_infinite} are \say{soft constraints} by nature, in that a small violation of the constraints is tolerable.
To handle the infinite dimensional constraints, \cref{eq:b_field_constraint_infinite}, we discretize the domain of the constraint into a uniform, $\ns\times \ntheta \times\nzeta$ grid. We then apply the magnetic field constraints at each of the $\ngrid = \ns \ntheta \nzeta$ grid points $\xb_i$,
\begin{equation}
\begin{split}
    &B_{-}^* - B(\xb_i) \le 0 \quad i=1,\ldots,\ngrid,
    \\
    &B(\xb_i) - B_{+}^*\le 0 \quad i=1,\ldots,\ngrid,
\end{split}
\label{eq:b_field_constraint_discrete}
\end{equation}
totaling $2\ngrid$ nonlinear simulation based constraints.

\subsection{Optimization objective}
\label{sec:objective}

Fast-ion optimization has two principle goals: minimizing the thermal energy lost from the system, and dispersing or concentrating the load of fast-ions on the plasma-facing components. We focus solely on the first goal of achieving excellent confinement of energy, noting that this also makes progress towards the second goal.

The confinement of fast-ions is often measured by the loss fraction, the fraction of particles lost within a terminal time $\tmax$. 
While the loss fraction measures particle confinement, it does not reflect the fact that particles lost quickly, with energy of nearly 3.5 MeV, contribute more to the heat flux on plasma-facing components and detract more from plasma self-heating than particles lost at late times, which have slowed substantially.
If collisions were included in the particle tracing calculations, the energy loss fraction could be computed straightforwardly.
However particle tracing is often done without collisions, because it is easier to implement and because efficient algorithms can be applied in the collisionless case \cite{albert2020accelerated,albert2020symplectic}.
Therefore, here we describe a physically motivated objective function that places greater weight on minimizing prompt losses within collisionless calculations.

Fusion-produced alpha particles primarily experience collisions with electrons during which they deposit most of their energy. 
% This follows from the fact that the slowing-down collision frequency \cite{NRLPlasma2019} for test alpha particles with background electrons is higher than the frequency for alphas to slow down on the ions as long as the alpha energy exceeds $\sim 50 T_e$ \cite[page 40]{helander2005collisional}. 
This follows from the fact that the slowing-down collision frequency \cite{NRLPlasma2019} for alpha particles with background electrons is higher than the slowing-down collision frequency with ions as long as the alpha energy exceeds \response{a critical energy of} $\sim 50 T_e$ \cite[page 40]{helander2005collisional}. 
If reactor temperatures satisfy $ T_e \le 16$ keV, then collisions with electrons dominate until the alphas have slowed to $\le 0.8$ MeV.
This process can be described by
\begin{equation}
    \frac{d v}{dt} = -\nu_s^{\alpha/e} v,
    \label{eq:slowing_down}
\end{equation}
where $\nu_s^{\alpha/e}$ is the alpha-electron slowing-down collision frequency, which is approximately independent of alpha energy \cite{NRLPlasma2019}.
\response{Note that slowing down has a higher frequency than pitch angle scattering or energy diffusion until the alphas have slowed to the critical energy. It is a reasonable approximation to neglect these latter two processes here since we are primarily concerned with the loss of alpha energy. The slowing-down frequency}
% In reality, 
$\nu_s^{\alpha/e}$ will vary with time as the particle traverses regions of different density and temperature.
We neglect this complexity treating $\nu_s^{\alpha/e}$ as approximately constant, in which case the solution of \cref{eq:slowing_down} becomes
\begin{equation}
    v(t) = v(0) e^{- \nu_s^{\alpha/e} t }.
    \label{eq:energy_model}
\end{equation}
The slowing-down time, $1 /  \nu_s^{\alpha/e}$, is typically on the order of $100\text{ms}$ for plausible reactor parameters.
Assuming an initial energy of 3.5 MeV, the energy lost associated with an alpha particle lost at time $\Tc$ is $3.5e^{-2 \nu_s^{\alpha/e} \Tc}\text{MeV}$. 

\begin{figure}[tbh!]
\includegraphics[scale=0.5]{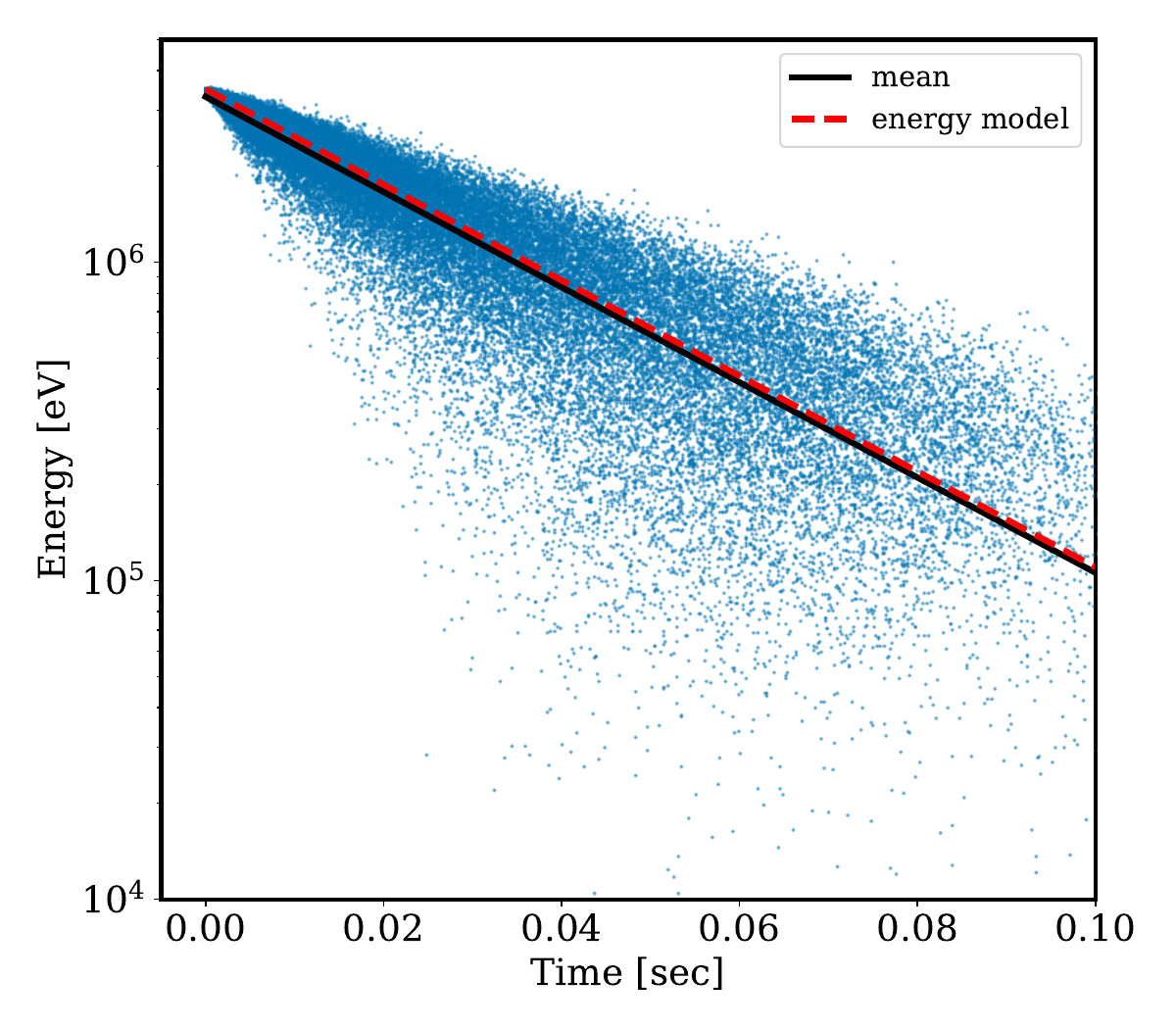}
\centering
\caption{Energy of alpha particles at the time they are lost. Data points were generated by tracing particles \textit{with collisions} from 10 configurations: NCSX, ARIES-CS, a QA from NYU, CFQS, a QH from IPP, a QA from IPP, HSX, Wistell-A, LHD, and W7-X. $20,000$ particles were traced for each configuration. The solid black line indicates the regressed mean of the data, and the dashed red line is the energy decay model $3.5\exp(-2t\nu_{s}^{\alpha/e})$ where the slowing-down time $1/\nu_s^{\alpha/e} \approx 0.057$sec was computed analytically using the volume averaged density and temperature \cite{NRLPlasma2019}. The energy model very closely matches the mean particle energy.}
\label{fig:objective_validation}
\end{figure}

In \Cref{fig:objective_validation} we see that the energy decay model \cref{eq:energy_model} is almost identical to the mean energy of alpha particles at any given time. Data for \Cref{fig:objective_validation} was generated using \textit{collisional tracing} in the ANTS code \cite{drevlak2014}. $20,000$ particles were traced from each of 10 configurations: the National Compact Stellarator eXperiment (NCSX) \cite{zarnstorff2001physics}, Advanced Research Innovation and Evaluation Study - Compact Stellarator (ARIES-CS) \cite{najmabadi2008aries}, a quasi-axisymmetric (QA) stellarator developed at New York University (NYU) \cite{garabedian2008three,garabedian2009design}, the Chinese First Quasi-axisymmetric Stellarator (CFQS) \cite{liu2018magnetic}, a quasi-helically (QH) symmetric stellarator developed at the Max Planck Institute for Plasma Physics (IPP) \cite{nuhrenberg1988quasi}, a QA stellarator developed at IPP \cite{henneberg2019properties}, the Helically Symmetric eXperiment (HSX) \cite{anderson1995helically}, Wistell-A \cite{bader2020advancing}, the Large Helical Device (LHD) \cite{iiyoshi1999overview}, and the Wendelstein 7-X (W7-X) \cite{klinger2016performance}. Scattered are alpha particle energies at they moment they are lost. The mean of the particle energies (solid black line) is shown against the energy model \cref{eq:energy_model} (dashed red line). The accuracy of the energy model in predicting the mean energy justifies its use as an optimization objective.

We take the expectation of this energy measure to compute our optimization objective, replacing $\nu_s^{\alpha/e}$ by the inverse of the fixed tracing time $\tmax$:
\begin{equation}
    \Jenergy(\xb,\vpar,\wb) = 3.5e^{-2\Tc(\xb,\vpar,\wb) / \tmax}
    \label{eq:Jenergy}
\end{equation}
\response{Using $\tmax$ instead of $1/\nu_s^{\alpha/e}$ amplifies the difference between configurations with similar energy losses, exposing the variation to optimization methods.} We write the confinement time as $\Tc(\xb,\vpar,\wb)$ to explicitly denote its dependence on the initial particle position, parallel velocity, and decision variables. For a particle that is lost at time $t$ the confinement time is calculated as $\Tc = \min\{t,\tmax\}$.
To compute our optimization objective, the expected energy lost, $\Jc(\wb) = \Ebb[\Jenergy(\xb,\vpar,\wb)]$ we integrate  $\Jenergy$ against the distribution $f(\xb,\vpar)$ of initial particle positions and parallel velocities, 
\begin{equation}
        \Jc(\wb) := \int_{\xb}\int_{\vpar}  3.5e^{-2\Tc(\xb,\vpar,\wb) / \tmax} \,\, f(\xb,\vpar) \, \, d\vpar \, d\xb .
    \label{eq:objective_integral}
\end{equation}
In \Cref{sec:objective_approximation} we discuss three possible methods of computing this integral, by Monte Carlo (MC), by Simpson's rule, and by Quasi-Monte Carlo (QMC) \cite{lemieux2009MC}. 

As a simple alternative to this objective we can also minimize the energy lost from particles born on a single flux surface. This has the advantage of reducing the dimension of the objective computation. Hence we define the surface objective as
\begin{equation}
        \Jc_{s}(\wb) := \int_{\theta,\zeta}\int_{\vpar}  3.5e^{-2\Tc(\xb,\vpar,\wb) / \tmax} \,\, f(\theta,\zeta,\vpar | s) \, \, d\vpar \, d\theta \, d\zeta .
    \label{eq:surface_objective_integral}
\end{equation}

Previous stellarator designs which leveraged optimization of empirical alpha particle losses, ARIES-CS and a design by Gori et. al., used the expected value of confinement time, and the conditional expectation of the confinement time over particles which bounce as optimization objectives. In this study, we opt to use the energy loss objective $\Jc$ rather than mean confinement time due to the interpretation as energy. However, the mean confinement time and $\Jc$ may be related through Jensen's inequality,
\begin{equation}
    \Jenergy(\Ebb[\Tc]) \le \Ebb[\Jenergy(\Tc)] = \Jc(\wb).
\end{equation}
By a straightforward computation, $\Ebb[\Tc] \le -\frac{\tmax}{2}\ln(\frac{\Jc(\wb)}{3.5})$.
Hence maximizing the mean confinement time should reduce $\Jc$ and similarly minimizing $\Jc$ should increase the mean confinement time. The set of local minima for these two objectives is not in general the same. However, if there exist configurations with $0\%$ losses, then the objectives share the set of global minimizers.

\section{Numerical computation of objective}
\label{sec:objective_approximation}

Monte Carlo quadrature and deterministic numerical quadrature methods can be used to approximate the integral \cref{eq:objective_integral}. Whether spawned on a mesh or randomly according to some distribution, particles with initial position and parallel velocity $(\xb_i,(\vpar)_i)$ are traced through time until breaching the last closed flux surface, $s=1$, at some time $t \le \tmax$, or until the terminal tracing time is reached $t=\tmax$. The confinement time is calculated as $\Tc = \min\{t,\tmax\}$, which can be converted to the approximate energy lost due to potential particle ejection via \cref{eq:Jenergy}. Quadrature methods combine the integrand values computed from evaluation points as a weighted sum,  
\begin{equation}
    \Jc(\wb) \approx \sum_{i=1}^N \omega_i\Jenergy(\xb_i,(\vpar)_i, \wb)f(\xb_i,(\vpar)_i, \wb)
    \label{eq:objective_quadrature}
\end{equation}
where the weights $\{\omega_i\}_{i=1}^N$ and nodes $\{\xb_i,(\vpar)_i\}_{i=1}^N$ are determined by the quadrature method. We briefly explore three different methods for approximating our objectives: MC, QMC, and Simpson's rule \cite{burden2015numerical}.

MC quadrature samples $N$ nodes randomly from some density $\{\xb_i,(\vpar)_i\}_{i=1}^N \sim g(\xb,\vpar)$ and approximates the integral via \cref{eq:objective_quadrature} with weights $\omega_i = (g(\xb_i,(\vpar)_i)N)^{-1}$. In our setting, $f_{\theta,\zeta}(\theta,\zeta |s)$ varies depending on the MHD equilibrium computed from $\wb$. To simplify the sampling procedure, we opt to sample $\theta$ and $\zeta$ from a uniform distribution. Hence, initial particle positions and velocities are sampled from
\begin{equation}
    g(s,\theta,\zeta,\vpar) := f_s(s)f_{\vpar}(\vpar)\nfp/4\pi^2.
    \label{eq:sample_density}
\end{equation}
The standard deviation, and hence convergence rate, of the MC estimator is $\sigma/\sqrt{N}$ where $\sigma$ is the standard deviation of $\Jenergy f/g$. On one hand MC is slow to deliver accurate estimates, but on the other hand it does not rely on smoothness assumptions to achieve its converge rate, unlike Simpson's rule. 

When used in the optimization loop, Monte Carlo methods can be applied in two ways: by regenerating the samples $\{(\xb_i,(\vpar)_i\}_{i=1}^N$ at each iteration, or by generating the samples once and holding them fixed throughout the optimization. We denote the former method as generic MC. The later method is known as the Sample Average Approximation method (SAA) \cite{shapiro2001monte}. A great benefit of using SAA is that it forms deterministic optimization problems which can solved by the any conventional optimization method. The principal drawback of SAA is the slight bias it incurs in the solution, similar to quadrature methods. When using generic MC to compute the optimization objective, stochastic optimization methods must be used to solve the optimization problem. Stochastic solvers tend to converge slowly, but arrive at unbiased solutions. 

Quasi-Monte Carlo methods are a deterministic analog of MC methods. Similar to MC they approximate integrals as sample averages. However, the points used in the sample average are not truly random, rather they are \textit{low discrepancy sequences}, approximately random sequences. Quasi-Monte Carlo methods boast a convergence rate of $O(1/N)$, when using $N$ points in the approximation, which is an impressive improvement over MC and SAA. The constant in the convergence rate depends on the \textit{total variation} of the integrand, a measure of it's rate of change, rather than it's variance. Since the integrand in \cref{eq:objective_integral} depends on the confinement time, which is non-smooth, and perhaps even discontinuous in $\xb$ and $\vpar$, the total variation of the integrand is large, and so QMC may not outperform MC until the number of samples is large.

\begin{figure}[tbh!]
\includegraphics[scale=0.4]{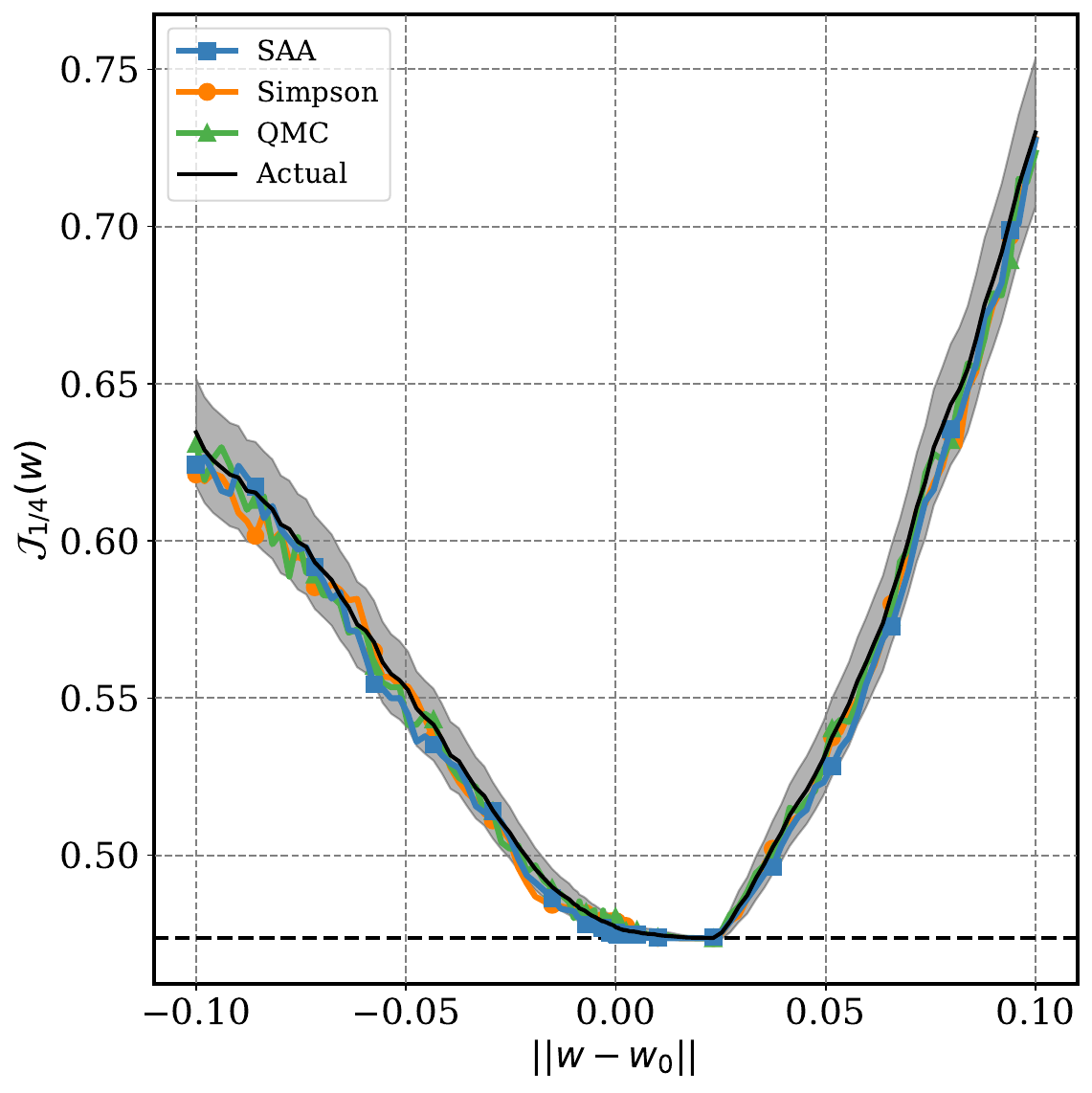}
\includegraphics[scale=0.4]{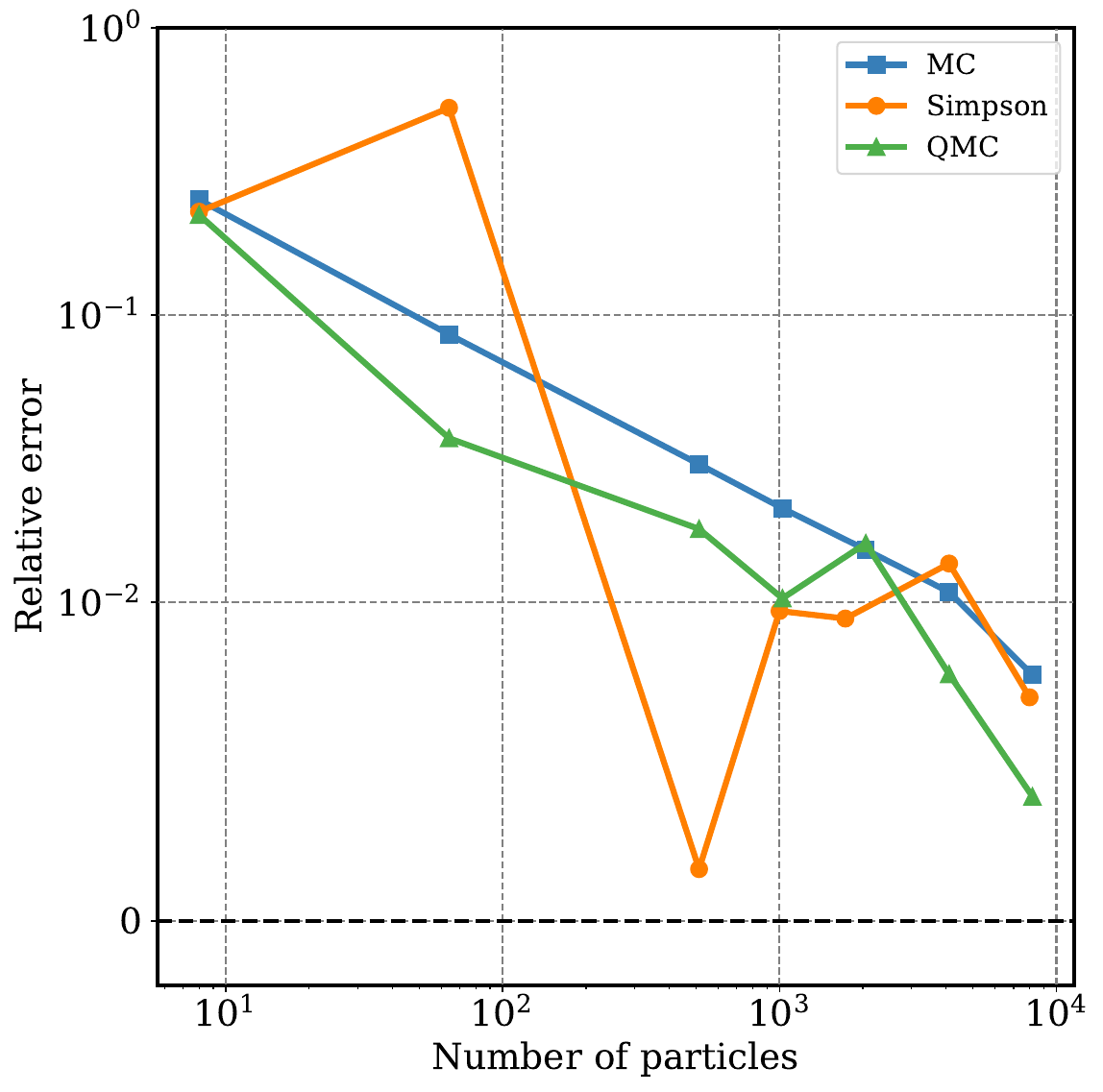}
\centering
\caption{(Left) Approximations of the objective function $\Jc_{1/4}$ using Simpson's rule, QMC, SAA, and MC across a one dimensional slice of space, around a point $\wb_0$. The curves were computed by tracing $4096$ particles per point, with particles on a $16^3$ mesh for Simpson's rule. The shaded region is the $95\%$ confidence interval for the objective value computation with MC. The black line (Actual) represents the actual value of the objective, and is computed using MC with $32,000$ samples. (Right) Relative error of MC, QMC, and Simpson's rule in computing the objective at a single point. \response{The relative error was computed against the QMC estimate with $2^{15}$ particles.} The MC curve represents the expected relative error of the MC estimator given the sample size, and was computed by \response{the statistical method} of bootstrapping \cite{james2013introduction}.}
\label{fig:quadrature_comparison}
\end{figure}

Simpson's rule uses quadratic interpolation of a function on a mesh to approximate the function's integral. High order quadrature methods, like Simpson's rule, achieve high-order convergence rates when the integrand can be well-approximated by a low-degree polynomial. However, since particle confinement times may jump chaotically under small perturbations in $\xb,\vpar$, Simpson's rule and other high-order quadrature schemes are not expected to achieve a high-order convergence rate. 

In \Cref{fig:quadrature_comparison}, we compare the approximation quality of four methods of computing the $\Jc_{1/4}$: generic MC, SAA, Simpson's rule, and QMC. \Cref{fig:quadrature_comparison} (right) shows the relative error of MC, Simpson's rule and QMC in approximating the objective $\Jc_{1/4}$ at a single point. Given the limits on sample size requirements, MC achieves similar accuracy to  Simpson's rule and QMC. QMC performs slightly better than MC, but does not reliably do so at the sample sizes shown. \Cref{fig:quadrature_comparison} (left) shows the objective approximations over a one dimensional slice of space near an arbitrary configuration $\wb_0$. Spatially, we find that SAA provides a smooth approximation to the objective, which is beneficial for optimization. For this reason we use SAA to compute the objectives in the numerical experiments. Unfortunately, due to the extraordinarily high standard deviation of the confinement times typically $>2000$ points are required to reduce the noise in the objective enough so that it can be tractably minimized by an optimization routine. The standard deviation of the confinement times is often of the same order of magnitude as the mean, though it decreases as the loss fraction decays to zero. In future work, variance reduction techniques \cite{law2023meta,law2022accelerating,hammersley2013monte} should be used to improve the accuracy of the objective computation and reduce the computational burden associated with tracing particles.

\section{Numerical results}
\label{sec:numerical}

\begin{figure}[tbh!]
\includegraphics[scale=0.21]{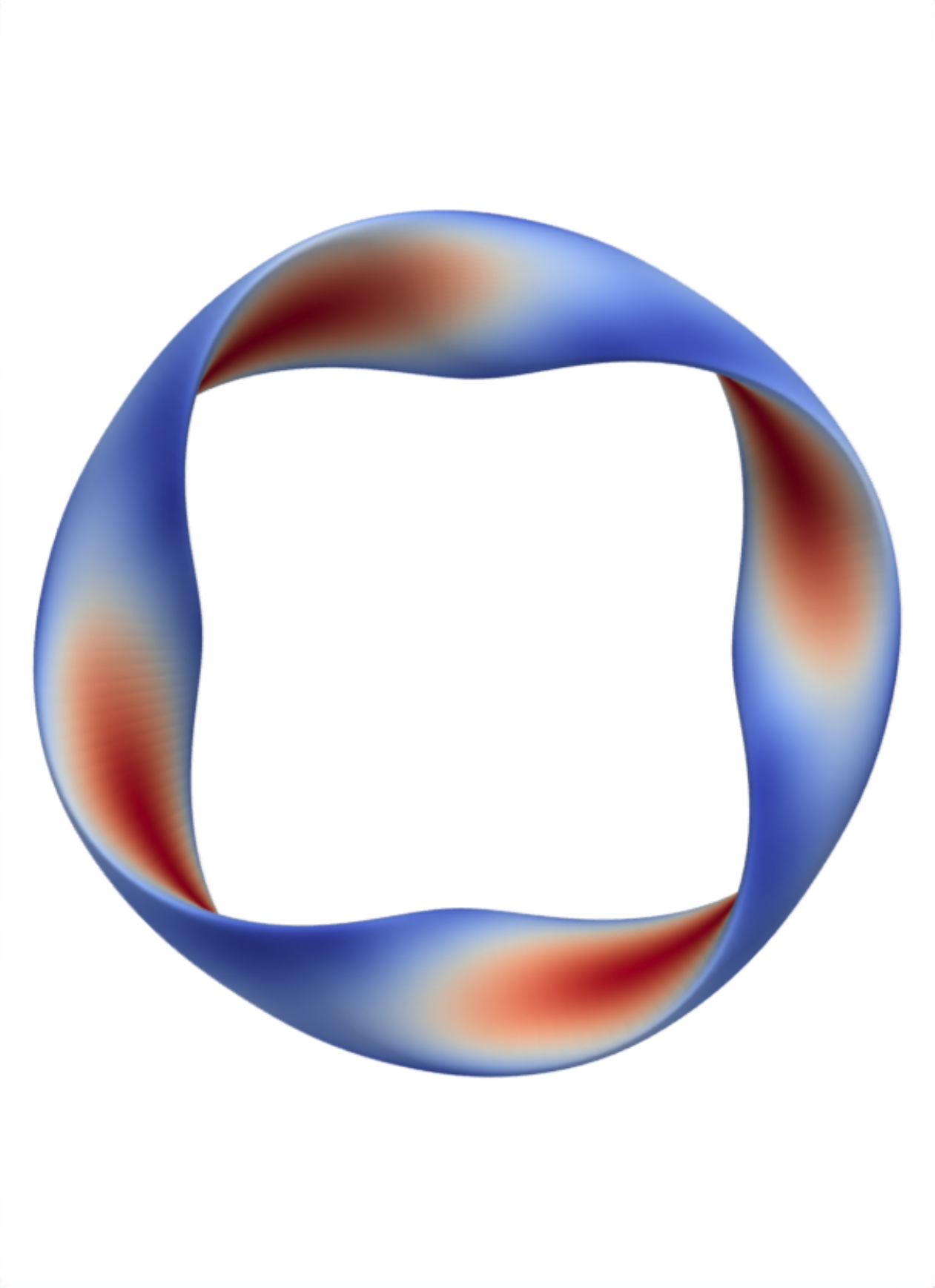}
\hspace{15pt}
\includegraphics[scale=0.21]{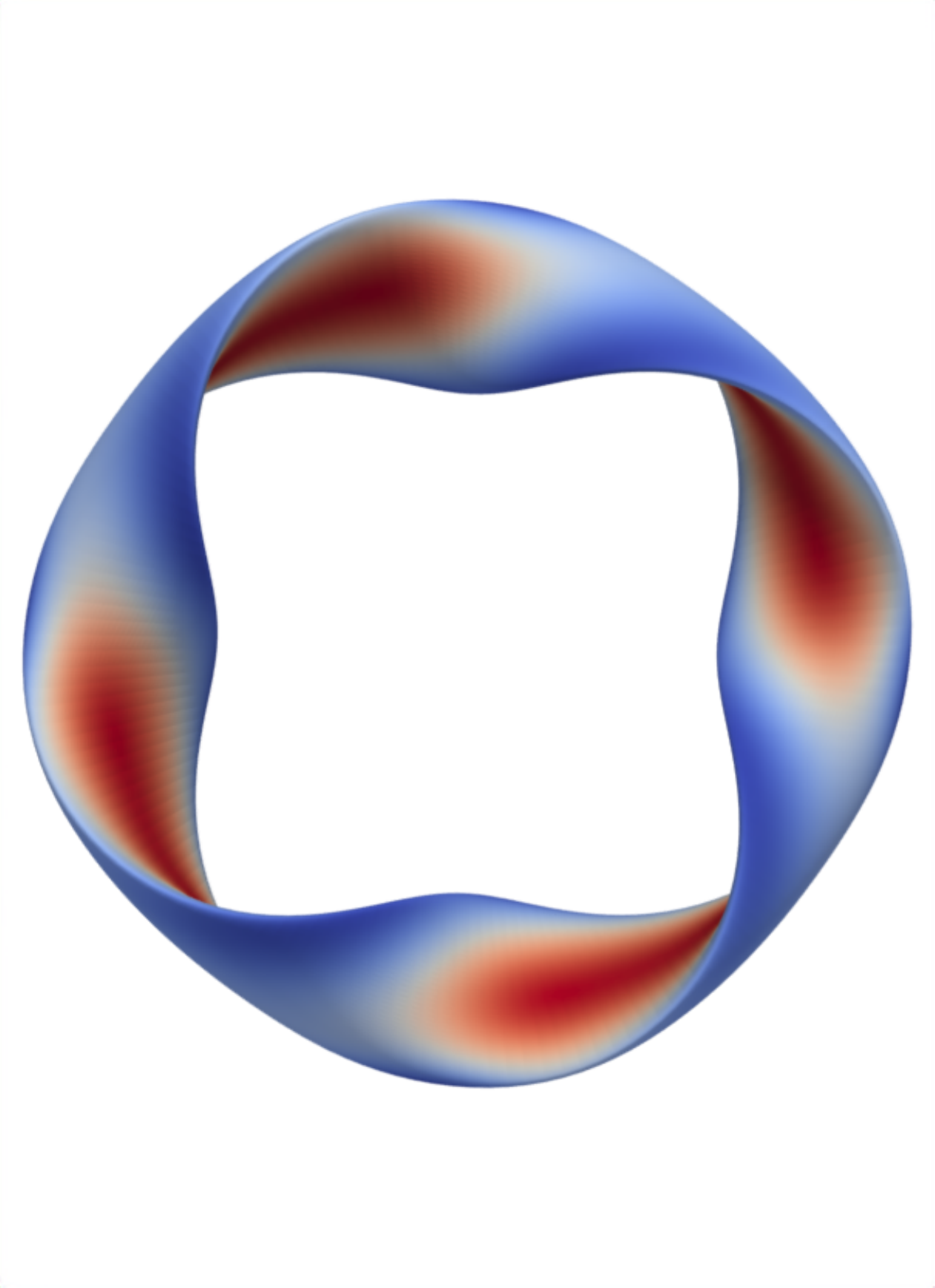}
\includegraphics[scale=0.3]{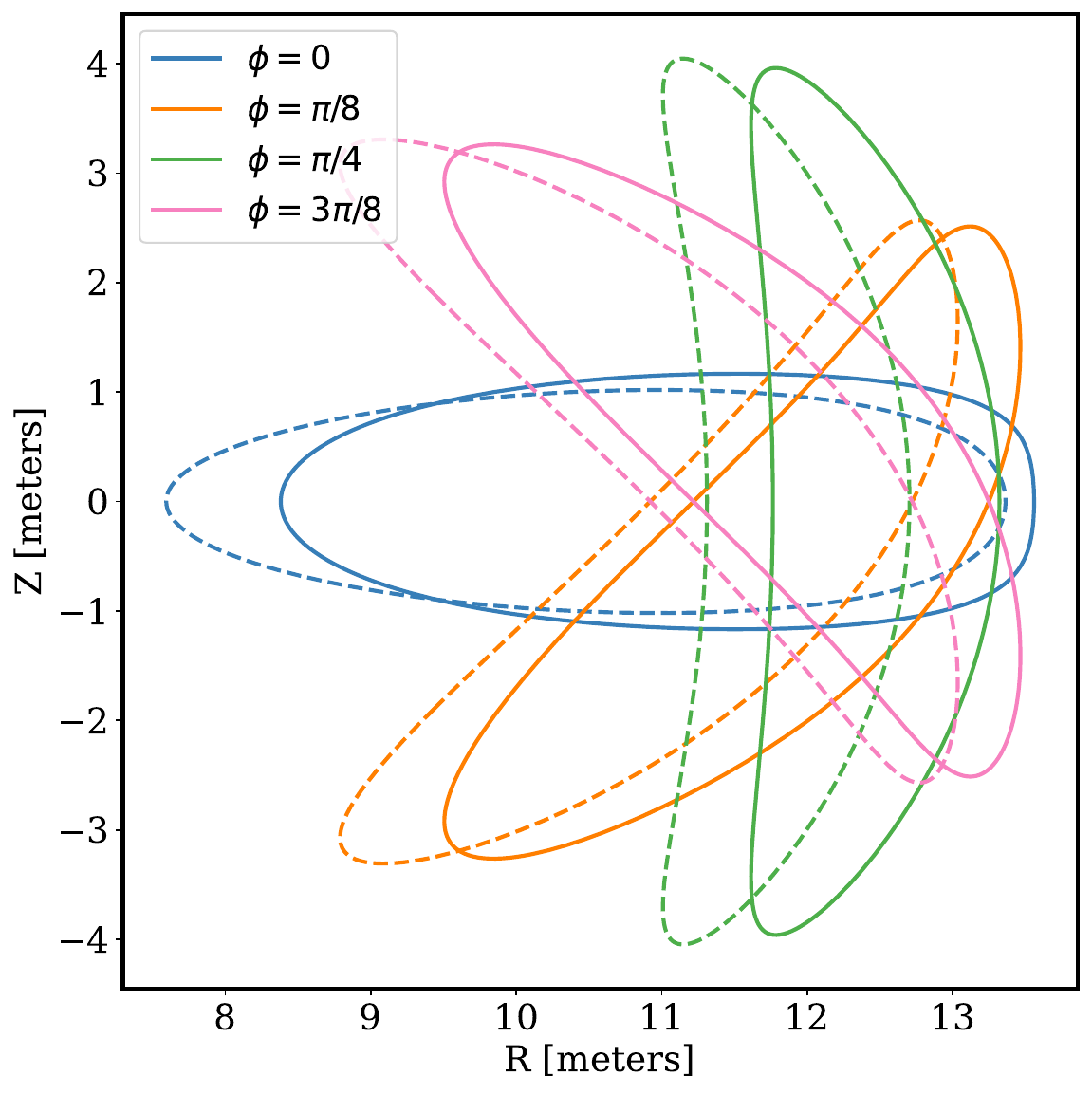}
\centering
\caption{Three dimensional views of configuration \textbf{A} (left) and configuration \textbf{B} (middle). Cross sections of configuration \textbf{A} (solid) and \textbf{B} (dashed) at four cylindrical angles $\phi$ across a field period (right).}
\label{fig:3d_configs}
\end{figure}

In this section we explore numerical solutions of \cref{eq:main}. We show physical properties of two, four field period vacuum configurations: configuration \textbf{A} was optimized using the surface initialization loss $\Jc_{1/4}$, and configuration \textbf{B} optimized using the volumetric initialization loss $\Jc$. 
We find that minimizers of $\Jc_{1/4}$ also perform well under $\Jc$, and that quasi-symmetry need not be satisfied for good confinement. Furthermore, we analyze the local relationship of particle losses with a quasi-symmetry metric, finding that reducing the violation of quasi-symmetry can increase particle losses. While our numerical solutions are vacuum configurations, the optimization model and numerical methods can be applied to finite-$\beta$ configurations as well. Due to the computational expense of repeated particle tracing, our configurations were optimized with a terminal trace time of $\tmax=10$ms. A three dimensional view of the configurations is shown in \cref{fig:3d_configs}. The data that support the findings of this study are openly available at 
%\url{https://github.com/mishapadidar/alpha_particle_opt}
\url{https://doi.org/10.5281/zenodo.7838152}.

\subsection{Methods}
\label{sec:methods}

Initial experimentation demonstrated that the optimization landscape contains many local solutions. To this end it was useful to search the optimization space by generating a host of initial points with varied rotational transform values, $\iota$. Starting points for the fast-ion optimization were generated by solving the optimization problem,
\begin{align}
    \minimize_{\wb} \ \  (A-A^*)^2 + (\iota-\iota^*)^2  + \sum_{i=1}^{\ngrid} \max(B(\xb_i) - B_+^*,0)^2 + \max(B_-^* - B(\xb_i),0)^2 ,
    \label{eq:phase_one}
\end{align}
in \code{SIMSOPT} using concurrent function evaluations to compute forward difference gradients, and the default solver in Scipy's least-squares optimization routine \cite{virtanen2020scipy}. \response{The optimization of \cref{eq:phase_one} was initialized from a purely toroidal boundary shape with aspect ratio $A^*$.} Optimal solutions were found to the problem within $5\%$ error for each target rotational transform $\iota^*$. The decision variables were characterized by $\nmode=1$, i.e $\nw=8$. \response{Configuration \textbf{A} and \textbf{B} were initialized from solutions of \cref{eq:phase_one} with $\iota^* = 0.89, 1.04$, respectively.}

The fast-ion optimization was initialized from the solutions of \cref{eq:phase_one}. The magnetic field bound constraints \cref{eq:b_field_constraint_discrete} were treated with a quadratic penalty method with penalty weights all equal to one,
 \begin{align}
    \minimize_{\wb} \ \  \Jc_{\text{penalty}}(\wb) := \Jc(\wb) +  \sum_{i=1}^{\ngrid} \max(B(\xb_i) - B_+^*,0)^2 + \max(B_-^* - B(\xb_i),0)^2, 
    \label{eq:penalty_method}
\end{align}
with an analogous form for using the surface objective $\Jc_{1/4}$. The particle loss objective $\Jc(\wb)$ was computed using SAA, since it provided a reasonably smooth approximation of the objective. The penalty method was used because the field strength constraints are \say{soft constraints} --- they do not need to be satisfied exactly. Powell's BOBYQA algorithm \cite{powell2009bobyqa} within the  Python package PDFO \cite{ragonneau2021pdfo} was used to solve \cref{eq:penalty_method}. BOBYQA is a derivative-free trust region method that uses local quadratic approximations of the objective to make progress towards a minimum. BOBYQA performed particularly well in this problem due to its ability to handle computational noise and use samples efficiently \cite{cartis2019improving}.

Empirically we find that the efficiency of the optimization with terminal time $\tmax$ can be substantially improved by warm-starting the optimization from a solution with near-zero losses at a shorter value of $\tmax$, say $\tmax/10$. The optimization up to the terminal time $\tmax=10$ms was performed solving a sequence of optimization problems, where at each step $\tmax$ and the number of Fourier modes were increased: $(\tmax,\nmode) =$(0.1\text{ms}, 1), (1\text{ms}, 1), (1\text{ms}, 2), (10\text{ms}, 2), (10\text{ms}, 3). For $\tmax=0.1,1,10$ms we use $10^4,7^4,6^4$ particles, respectively, and $8,48,48$ MPI processes to trace particles. Particles were traced until reaching the terminal tracing time of $\tmax$, or until the particle reached the $s=1$ flux surface or $s=0.01$ flux surface. Particles reaching the $s=1$ flux surface were deemed lost, while particles reaching the $s=0.01$ flux surface were deemed to be confined to the terminal time $\tmax$. The $s=0.01$ stopping criteria is currently required as part of the tracing code in \code{SIMSOPT}, but should not be used in future work.

\response{During the optimization, \code{VMEC} was run with 50 surfaces, and MPOL=NTOR=7. The Boozer transformation was performed with 16 poloidal and toroidal modes.}%Cubic splines were used to interpolate the field.

\begin{table}[htb!]
\def\arraystretch{2}
\centering
\begin{tabular}{|c|c|c|c|c|c|c|}
\hline
Config. & Aspect Ratio & Mirror ratio & Mean $\iota$ & Volume loss fraction & $s=1/4$ loss fraction & \response{$\epsilon_{eff}$}
\\
\hline
\textbf{A} &6.67 &1.33 &0.856 &0.022 &0.0046 & \response{$<1.2\%$}
\\
\hline
\textbf{B} &6.61 &1.32 &1.023 & 0.0215 &0.0094 & \response{$<0.82\%$}
\\
\hline
\end{tabular}
\caption{Properties of configurations \textbf{A} and \textbf{B}. Loss fractions were computed by tracing 10,000 particles to $\tmax=10$ms.
}
\label{table:config_A_B}
%\vspace{5pt}
\end{table}

\subsection{Two solutions}
\label{sec:two_solutions}

We present two solutions found by solving \cref{eq:main}. Configuration $\textbf{A}$, with solution vector $\wb_{\textbf{A}}$, was found by minimizing the surface initialization objective $\Jc_{1/4}$ which measures the energy lost by particles born on the $s=0.25$ flux surface. Configuration $\textbf{B}$, with solution vector $\wb_{\textbf{B}}$, was found by minimizing the energy lost by particles born throughout the entire volume, i.e. objective $\Jc$. Properties of configuration \textbf{A} and \textbf{B} can be seen in \Cref{table:config_A_B}. All configurations presented in this section were scaled to same $1.7$m minor radius and $B_{0,0}(s=0) = 5.7$T field strength on the magnetic axis as the ARIES-CS configuration \cite{ku2008physics}. 

Configuration $\textbf{A}$ \textit{almost} reaches the global minimum value of $\Jc_{1/4}$, attaining an objective value of $\Jc_{1/4}(\wb_{\textbf{A}}) = 0.475$ and a loss fraction of $0.0046$ for particles born on the $s=0.25$ flux surface; a global minimum would have zero particle losses and $\Jc_{1/4} = 0.473$. 
% Configuration \textbf{A} still may be far from a global minimum in configuration space, since $\Jc_{1/4}$ is fairly flat around $\wb_\textbf{A}$.
Configuration $\textbf{A}$ also reports a low loss fraction for particles born, according to $f$, throughout the volume, $0.022$. Similarly, configuration \textbf{B} has a loss fraction of $0.0215$ for particles born throughout the volume and a loss fraction of $0.0094$ for particles born on the $s=0.25$ flux surface. While the two configurations were optimized for different objectives, both configurations show good performance in both objectives. Optimizing using the surface loss $\Jc_{1/4}$ reduces the dimension of the integral \cref{eq:surface_objective_integral} and potentially the variance of the objective. Since improvement in the two objectives is highly correlated, in future work the surface loss objective could be used in place of the volume loss objective $\Jc$, unless confinement times are largely dependent on the radial birth distribution due. 

Neither configuration \textbf{A} nor \textbf{B} has active constraints at the solution, and so the constraints do not limit the performance of the solutions. We do find however, that in general the constraints on the field strength are active throughout the optimization. Without constraining the field strength, the mirror ratio becomes unphysically large, the contours of $B$ close poloidally, and solutions become approximately Quasi-Isodynamic \cite{subbotin2006integrated}.

\begin{figure}[tbh!]
\includegraphics[scale=0.5]{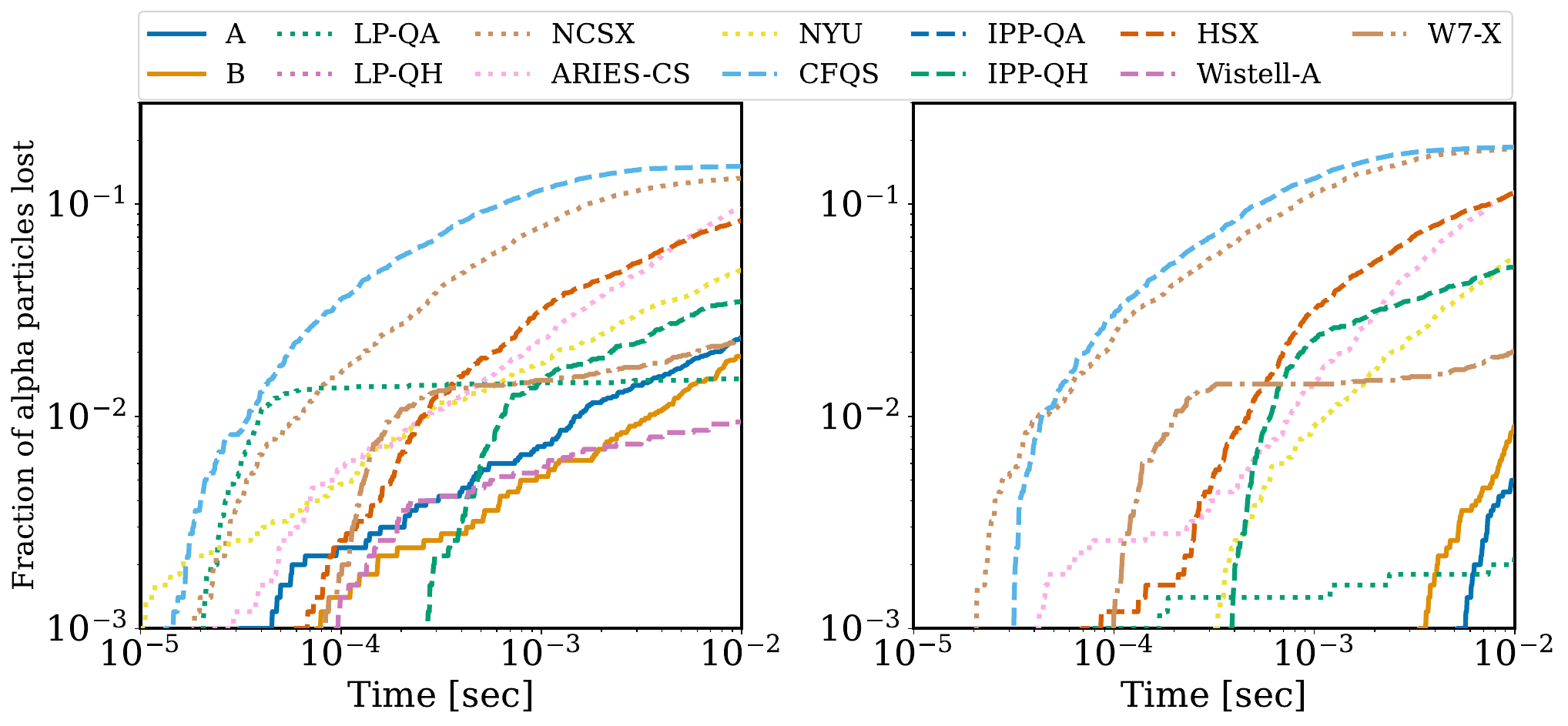}
\centering
\caption{Collisionless loss curves for $5000$ particles born throughout the volume with distribution $f$ (left) or born on the $s=0.25$ flux surface (right). Configurations were scaled to a $1.7$m minor radius and a $B_{0,0}(s=0)=5.7$T field strength on the magnetic axis, like the ARIES-CS reactor \cite{ku2008physics}. Wistell-A reported less than $0.1\%$ losses from the $s=0.25$ flux surface, and LP-QH reported no losses from the $s=0.25$ flux surface and a loss fraction of $0.0002$ for particles distributed throughout the volume.}
\label{fig:loss_profile}
\end{figure}

In \Cref{fig:loss_profile} we compare the alpha particle loss curves of configuration \textbf{A} and \textbf{B} to those of the stellarator configurations introduced in \Cref{fig:objective_validation}, as well as the QA and QH configurations from Landreman and Paul (labeled LP-QA and LP-QH)\cite{landreman2022magnetic}.
% NCSX, ARIES-CS, a quasi-axisymmetric (QA) stellarator developed at New York University (labeled NYU) \cite{garabedian2008three,garabedian2009design}, the Chinese First Quasi-axisymmetric Stellarator (CFQS) \cite{liu2018magnetic}, a quasi-helically (QH) symmetric stellarator developed at the Max Planck Institute for Plasma Physics \cite{nuhrenberg1988quasi}, the Helically Symmetric eXperiment (HSX) \cite{anderson1995helically}, Wistell-A \cite{bader2020advancing}, the QA and QH configurations from Landreman and Paul (labeled LP QA and LP QH)\cite{landreman2022magnetic}, and the Wendelstein 7X (W7-X) \cite{klinger2016performance}. 
To compute the curves, $5000$ particles born throughout the volume (left) or on the $s=0.25$ flux surface (right), were traced until the terminal time 10ms or until either they crossed the $s=1$ flux surface and were considered lost, or reached $s=0.01$ and were considered confined. Our configurations demonstrate good particle confinement up to the terminal time $\tmax=10$ms used in the optimization. Configuration \textbf{A} and \textbf{B} outperform all but LP-QA, LP-QH and Wistell-A in terms of particle losses from the $s=0.25$ flux surface, and are only outperformed by Wistell-A and LP-QH in terms of losses of volumetrically initialized particles. The lowest loss fraction from the $s=0.25$ flux surface, $0\%$, is that of LP-QH. The QS optimization problem posed by Landreman and Paul is computationally much less expensive to solve, and has a much smoother objective than $\Jc$ and $\Jc_{1/4}$, allowing for solutions to be refined to a much \response{finer resolution} with gradient-based optimization methods. 

\begin{figure}[tbh!]
\includegraphics[width=3.5in]{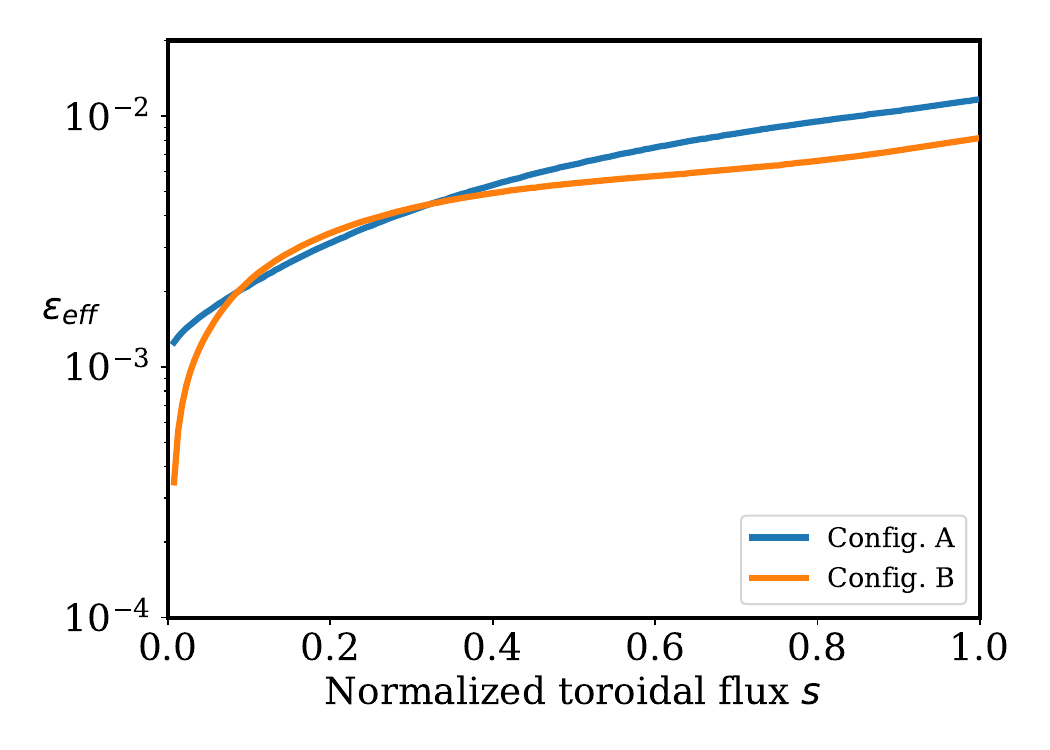}
\centering
\caption{\response{Thermal neoclassical transport metric $\epsilon_{eff}$ for the two optimized configurations.}}
\label{fig:epsilon_eff}
\end{figure}

\response{
It is noteworthy that the new optimized  configurations here not only have good confinement of energetic particles but also good neoclassical confinement of the thermal plasma.
The quantity $\epsilon_{eff}$ used as a measure of thermal neoclassical transport is plotted for the two configurations in \Cref{fig:epsilon_eff}.
For configuration \textbf{A}, $\epsilon_{eff} < 1.2\%$, and for configuration \textbf{B}, $\epsilon_{eff} < 0.82\%$, with lower values in the core in both cases.
These values are likely to be sufficiently small for a reactor \cite{alonso2022physics}.
Note that $\epsilon_{eff}$ was not included in the objective function. However since $\epsilon_{eff}$ is an average of the bounce-averaged radial guiding center drift, and the bounce-averaged radial drift is also the driver of alpha loss, it is not surprising that the optimization for alpha confinement here naturally leads to small $\epsilon_{eff}$ as a side-effect.
}

\subsection{Local analysis of Quasi-symmetry}
\label{sec:quasisymmetry_analysis}

\begin{figure}[tbh!]
\includegraphics[scale=0.4]{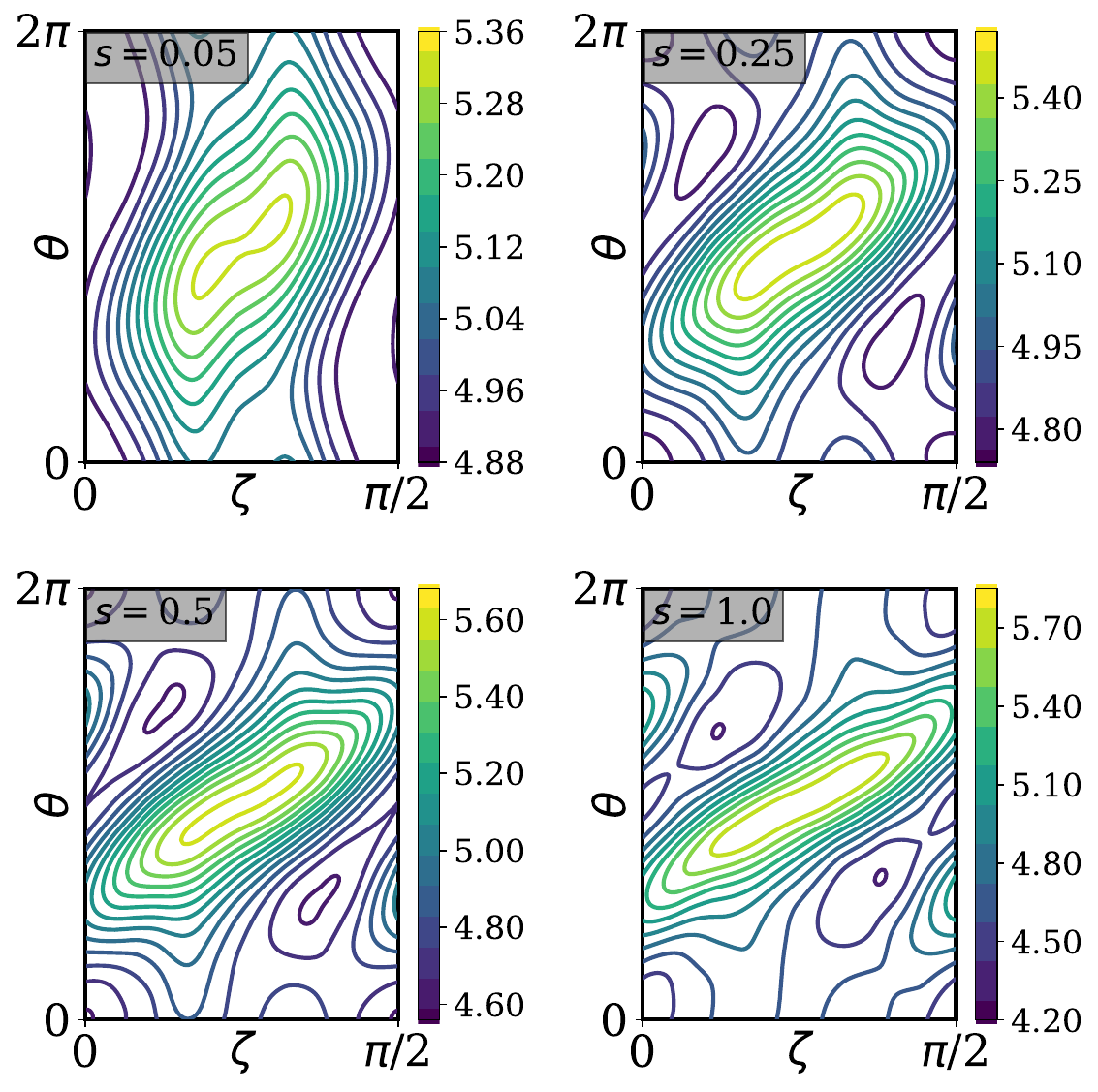}
\includegraphics[scale=0.4]{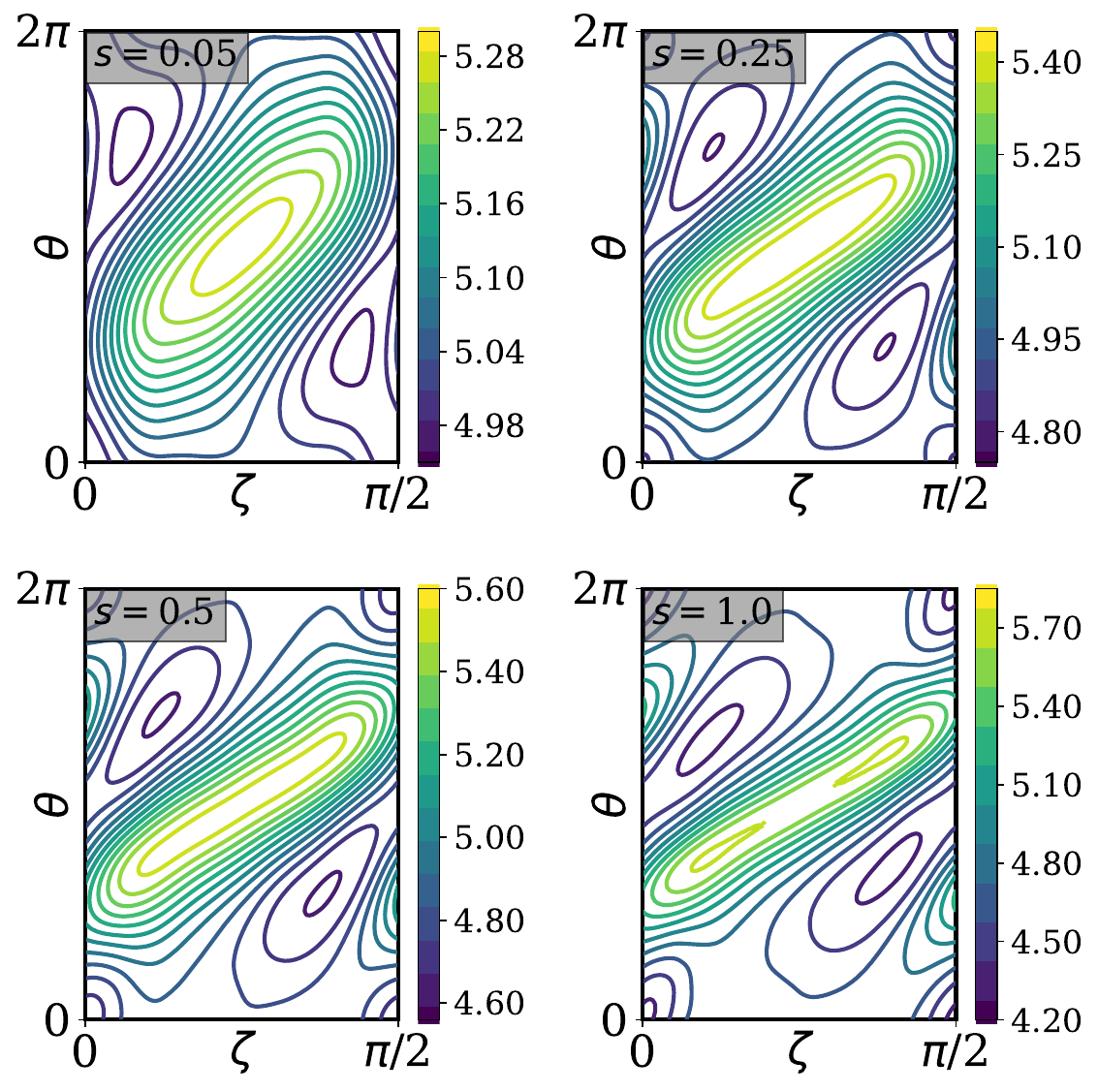}
\centering
\caption{Contours of $B$ in the Boozer coordinates on four flux surface $s=0.05,0.25,0.5,1.0$, for configuration \textbf{A} (left four) and configuration \textbf{B} (right four).}
\label{fig:B_contour}
\end{figure}

Neither configuration \textbf{A} nor configuration \textbf{B} are QS. This is seen most clearly in \Cref{fig:B_contour} by viewing the contours of the magnetic field strength in Boozer coordinates. QS fields have the representation $B = B(s,m\theta - n\zeta)$ for some numbers $m,n$ in Boozer coordinates, implying that the contours of $B$ are straight when viewed as a function of $\theta$ and $\zeta$ \cite{imbert2019introduction}. 
Near the magnetic axis only $m=1$ is possible \cite{cary1997helical}, and to preserve field period symmetry $n$ must be a multiple of the number of field periods, $n \in k\nfp$ for $k\in\mathbb{N}$. Quasi-axisymmetry occurs when $m=1,\, n=0$ and quasi-helical symmetry occurs when $m=1,\, n\neq 0$.

Exact quasi-symmetry is a sufficient condition for perfect confinement. In addition, Landreman and Paul \cite{landreman2022magnetic} showed that precisely quasi-symmetric configurations can have excellent confinement properties. However, in general it is not clear how particle confinement degrades when QS is broken, or how particle confinement improves as the violation of QS is reduced.

\begin{figure}[tbh!]
\includegraphics[scale=0.6]{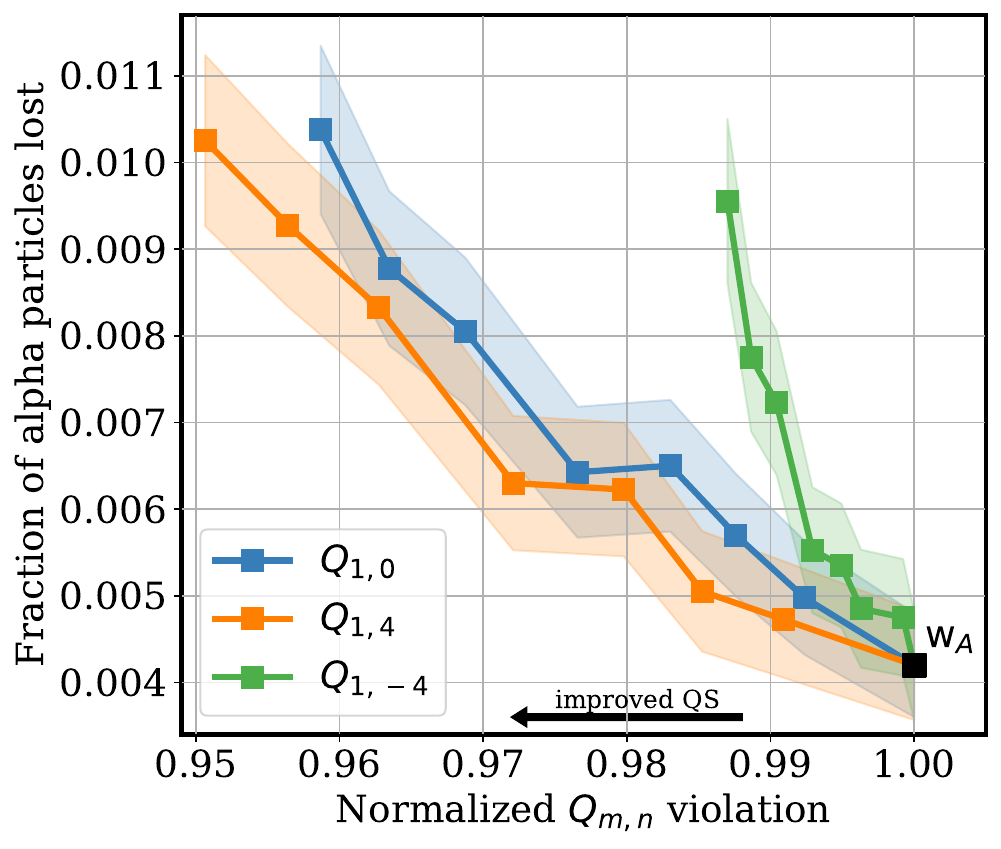}
\centering
\caption{Fraction of alpha particles lost as the violation of QS is reduced (right to left) along the line segment $\wb_{\textbf{A}} -\alpha\nabla Q_{m,n}(\wb_{\textbf{A}})$, for three different types of QS: $(m,n) = (1,0), (1,4), (1,-4)$. Reducing the violation of QS increases the fraction of lost particles. The shaded region indicates the $95\%$ confidence interval of the loss fraction.}
\label{fig:local_qs}
\end{figure}

To explore this relationship, we modify configuration \textbf{A} to reduce the violation of QS and examine how the corresponding particle losses are affected. The degree of $(m,n)$-quasi-symmetry of a configuration can be measured by the metric proposed in \cite{landreman2022magnetic}, which we denote $Q_{m,n}(\wb)$. Configuration \textbf{A} can be modified to have reduced violation of $(m,n)$-quasi-symmetry by moving the solution vector $\wb_{\textbf{A}}$ in the negative gradient direction of $Q_{m,n}$. For small step sizes $\alpha$, configurations with decision variables 
\begin{equation}
    \wb_{m,n}(\alpha) = \wb_{\textbf{A}} -\alpha\nabla Q_{m,n}(\wb_{\textbf{A}})
    \label{eq:step_in_qs_dir}
\end{equation}
will have a lower departure from QS than configuration \textbf{A}. As seen in \Cref{fig:local_qs}, as the violation of QS is reduced along this path, particle losses increase substantially, from approximately $0.43\%$ to approximately $1\%$, for all three types of QS considered $(m,n) = (1,0), (1,4), (1,-4)$. Locally, the violation of QS has an inverse relationship with confinement and so quasi-symmetric configurations may be isolated from non-quasi-symmetric configurations with low losses.

% \subsection{Loss distribution}
% \todo{Plot ideas: distribution of lost particles on boundary. Distribution of magnetic moment of lost particles.}

\section{Future work}
\label{sec:conclusion}

In the design of the ARIES-CS reactor, configurations with good alpha confinement were found by including alpha particle tracing calculations as part of an objective function within the optimization loop.
% In this study we explored stage one optimization of stellarator designs by directly relying on alpha particle simulations. 
In this study we have expanded upon this method, showing that it can be used to find configurations with fractional alpha losses. However, in it's current form, fast-ion optimization is computationally expensive, often taking multiple days to complete. Proxy metrics on the other hand, can be used to design stellarators in only a few hours on a computing cluster. To reduce the wall-clock-time of fast-ion optimization we propose three improvements: the use of variance reduction techniques to reduce the number of traced particles, symplectic particle tracing algorithms to improve the speed and accuracy of confinement time calculations, and multi-fidelity optimization methods to reduce the number of times particle tracing needs to be performed altogether. 

Law et. al. found in \cite{law2021accelerating,law2023meta} that combining variance reduction techniques such as importance sampling, control variates, and information reuse \cite{ng2014multifidelity} can reduce the number of particles that must be traced
% in order to resolve distribution statistics 
by a factor of 100. In addition, variance reduction techniques can be implemented relatively quickly making them an easily implementable addition to fast-ion optimization methods. The time spent tracing can also be reduced by improving orbit integration time. Albert et. al. \cite{albert2020accelerated,albert2020symplectic} showed that symplectic tracing algorithms can trace particle trajectories three times faster than adaptive integration algorithms, such as RK45, while maintaining the same statistical accuracy. Lastly, we propose using multi-fidelity optimization methods to reduce the number of expensive particle tracing simulations \cite{march2012provably,peherstorfer2018survey}. Multi-fidelity optimization methods for fast-ion optimization would rely on \say{low-fidelity models} of $\Jc$ to take reliable steps towards minima without performing many expensive particle tracing simulations. Low fidelity models of the energy loss objective could leverage particle tracing simulations with larger step sizes or simply be proxies, such as quasi-symmetry metrics.

In addition to improvements in optimization efficiency, there are improvements to be made in constructing objective functions. 
Thus far, particle tracing has only been used to measure confinement. 
% owever, particle tracing within the optimization loop has more potential use cases than confinement calculations, such as measuring the heat flux on plasma facing components. 
However, now that particle losses can be tractably reduced, the destructive effects of alphas on plasma-facing components is a central design consideration. A \say{wall-loading} objective function can either concentrate or disperse the alpha particle load on the wall, depending on engineering considerations. 

% Future work could consider finite beta optimization, and a fusion reaction rate that varies alongside the optimization. 

\section{Acknowledgments}
We thank Max Ruth, Shane Henderson, and Rogerio Jorge for their useful discussions. This work was supported by a grant from the Simons Foundation (No. 560651, D.B.).

\bibliography{references}
\bibliographystyle{plain}

% \appendix

\end{document}